\PassOptionsToPackage{unicode}{hyperref}
\PassOptionsToPackage{hyphens}{url}
\documentclass[
]{article}
\usepackage{hyperref}
\usepackage{natbib}
\usepackage{enumitem}
\usepackage{xcolor}
\usepackage{float}
\usepackage{amsmath,amssymb}
\usepackage{amsthm}
\theoremstyle{definition}
\newtheorem{definition}{Definition}
\theoremstyle{plain}
\newtheorem{proposition}{Proposition}
\newtheorem{corollary}{Corollary}
\usepackage{iftex}
\usepackage[T1]{fontenc}
\usepackage{lmodern} 
\IfFileExists{upquote.sty}{\usepackage{upquote}}{}
\IfFileExists{microtype.sty}{
  \usepackage[]{microtype}
  \UseMicrotypeSet[protrusion]{basicmath} 
}{}
\makeatletter
\@ifundefined{KOMAClassName}{
  \IfFileExists{parskip.sty}{%
    \usepackage{parskip}
  }{
    \setlength{\parindent}{0pt}
    \setlength{\parskip}{6pt plus 2pt minus 1pt}}
}{
  \KOMAoptions{parskip=half}}
\makeatother
\usepackage{longtable,booktabs,array}
\usepackage{calc} 
\usepackage{etoolbox}
\makeatletter
\patchcmd\longtable{\par}{\if@noskipsec\mbox{}\fi\par}{}{}
\makeatother
\IfFileExists{footnotehyper.sty}{\usepackage{footnotehyper}}{\usepackage{footnote}}
\makesavenoteenv{longtable}
\usepackage{graphicx}
\makeatletter
\newsavebox\pandoc@box
\newcommand*\pandocbounded[1]{
  \sbox\pandoc@box{#1}%
  \Gscale@div\@tempa{\textheight}{\dimexpr\ht\pandoc@box+\dp\pandoc@box\relax}%
  \Gscale@div\@tempb{\linewidth}{\wd\pandoc@box}%
  \ifdim\@tempb\p@<\@tempa\p@\let\@tempa\@tempb\fi
  \ifdim\@tempa\p@<\p@\scalebox{\@tempa}{\usebox\pandoc@box}%
  \else\usebox{\pandoc@box}%
  \fi%
}
\def\fps@figure{htbp}
\makeatother
\providecommand{\tightlist}{%
  \setlength{\itemsep}{0pt}\setlength{\parskip}{0pt}}
\usepackage[margin=1in]{geometry}
\usepackage{bookmark}
\IfFileExists{xurl.sty}{\usepackage{xurl}}{} 
\hypersetup{
  hidelinks,
  pdfcreator={LaTeX via pandoc}}

\title{\textbf{ACT, WAIT, or EXPERIMENT: A Causal Governance Framework for Retail Price Optimization Under Abstentions}}

\author{
  \textbf{Pedro Cadahia Delgado}\thanks{Corresponding author.} \\
  \small PhD, Universidad de Huelva, Spain \\
   \small\emph{Methodological study $\cdot$ Causal Pricing and Governance in Retail $\cdot$ 2026}
}

\date{} 

\begin{document}

\maketitle

\begin{abstract}
\noindent Price elasticities estimated from list-price variation in intermediated retail channels are confounded by pass-through, promotions and competitor moves, so an optimizer can act on estimates the evidence does not support. We propose a governance framework in which abstention is a primary output: an \textsc{act/wait} system built on Double Machine Learning, cost-shock instrumental contrasts and conformal prediction, whose identification-type \textsc{wait} flags a candidate for a designed pricing experiment. Under a local quadratic model, withholding beats acting exactly when the calibration slope of the true headroom on the recommended move is below one half; identification matters only through that slope. Two further contributions: seven tacit assumptions turned into executable contrasts, and an aggregation rule that cuts RMSE from 0.488 to 0.159. Resampled intervals undercover, and on Dominick's data one window's estimate forecasts later price changes poorly (calibration slope 0.22). In synthetic generators with known truth, the condition predicts the sign of the gain from acting in 37 of 38 groups. The system abstains wherever the condition fails, but also on 0.52 and 0.87 of identifiable presentations in the five-regime and independent generators, and loses more contribution than always acting in both base scenarios. It cuts the probability of a harmful move from 0.099 to 0.002 and from 0.357 to 0.036, though in the independent market mostly by acting rarely (a random rule at the same rate: 0.050). What transfers is a calibration-based audit of any abstention rule and a rule for the unit of publication. No commercial panel or pilot was used.

\vspace{0.6em}
\end{abstract}

\clearpage

\section{Introduction}\label{sec:intro}

A manufacturer selling through decentralized retail sets the list price and does not set the
price the consumer pays. The shopkeeper does, subject to local margin targets, menu costs and
rigid psychological price points \citep{levy2011price}. Any elasticity estimated from
list-price variation in this channel is therefore a composite: the consumer's response to the
shelf price convolved with the shopkeeper's pass-through of the list price to the shelf
\citep{besanko2005own}. The operational problem is nonetheless posed at the list price, since
that is the only lever the manufacturer holds: set monthly list prices, bounded by
$\pm 20\%$, so as to maximize contribution without sacrificing volume or share.

Naive regressions of volume on historical prices fail here because they confound the causal
slope with promotions, competitor moves, seasonality and nominal inflationary drift
\citep{villas1999endogeneity}. Since an optimizer is a causal intervention machine
\citep{pearl2009causality}, the binding problem in this channel is not optimization but the
isolation of the slope the optimizer consumes. The revenue-management literature and the
enterprise deployments it documents \citep{talluri2006theory, hormby2010marriott,
deng2023alibaba, llenas2026pepsico} largely take the observational price--demand relation as
sufficient for prescription. Decentralized Retail violates that premise through three structural
conditions: scarce list-price variation confounded with promotion, indirect and heterogeneous
pass-through, and shelf prices captured by audit rather than by transaction.
Section~\ref{sec:related-work} situates that gap against seven bodies of literature this work
draws on.

\paragraph{From estimation to decision.} Most of that literature, and most practice built on
it, treats the object of interest as a number: an elasticity to be estimated as precisely as
possible and handed to an optimizer. This paper treats it instead as an input to a decision
that is not always safe to make, and three questions replace ``what is the elasticity'' as its
central question.

\begin{enumerate}
  \item \textbf{When can a price change be executed on the estimate, and when should the system
  withhold a recommendation instead?} This is the \textbf{ACT/WAIT} question, resolved at the
  governance layer into structural or transient abstention, or an active candidate for
  EXPERIMENT, which yields the tri-part decision space of Act, Wait, or Experiment. Treating
  \emph{WAIT} as a first-class, reportable output, rather than as evidence of a system that
  failed to produce a number, is the paper's first contribution
  (Section~\ref{subsec:decision-layer}); Section~\ref{subsec:when-abstain} states when withholding
  beats acting and Section~\ref{subsec:condition-check} checks that statement on the records of
  the validation run.
  \item \textbf{What conditions, on the data and on the estimator, have to be made explicit and
  checked before an ACT verdict is trusted?} Converting assumptions such systems ordinarily leave
  tacit into executable contrasts, four executed by the system and three run as validation audits
  against a known truth, together with a declared account of which adapt an existing econometric
  tool and which are original to this setting, is the second contribution
  (Section~\ref{subsec:tacit}).
  \item \textbf{At what level of commercial aggregation, presentation, brand or category, is a
  decision legitimate when the evidence does not support it at the level the business would
  prefer to act on?} Showing when and how far the unit of decision should move is the third
  contribution (Section~\ref{subsec:unit-of-identification}).
\end{enumerate}

This paper reports a system built to answer those questions in one specific setting and, more consequentially, the validation run that measured whether it does. The system chains Double Machine Learning \citep{chernozhukov2018double}, a cost-shock instrumental contrast, hierarchical pooling, conformal predictive intervals, and a decision layer that abstains when identification is unviable. It is validated against two synthetic generators with known ground truth under thirty engineering-verification rules registered before the run (Section~\ref{subsec:validation-design}); the cost of its abstentions is then measured again in a third generator, written independently, and its estimates are checked against real price changes in public scanner data. Ten of the thirty rules pass, nineteen fail, and one does not apply. One failure is a cost this paper measures rather than defends: on its full decision function the system abstains on more than half of the presentations that are identifiable by construction (Section~\ref{subsec:funnel}), and on $0.87$ of them in a market generator written independently (Section~\ref{subsec:other-market}).

\paragraph{What a reader can take from this paper.} The validation returns more negative than positive verdicts on the system, deliberately: what transfers is the way of judging, not the system. Four things carry over to any pricing system that can abstain. (i)~An abstention rule earns something only through the covariance between acting and the gains from acting (Proposition~\ref{prop:value}), and acting beats holding exactly when the calibration slope of the true headroom on the recommended move exceeds one half (Proposition~\ref{prop:headroom}), a condition that predicts the sign of the gain in $37$ of $38$ groups. (ii)~Identifiability does not separate the groups where acting pays (Corollary~\ref{cor:classes}), so a gate should be judged by the calibration of what it lets through. (iii)~Publish at the unit the evidence supports: design bias averages at the square-root rate across independently priced presentations (RMSE $0.488\to0.159$ at the category level; Section~\ref{subsec:unit-of-identification}), but a resampled interval inside one realized design cannot see the dispersion across designs, so aggregation does not make it honest and some presentations are correctly answered with WAIT (Section~\ref{subsec:impossibility}). (iv)~Seven assumptions that a competent build leaves tacit become contrasts cheaper than the failures they detect. A reader should not conclude that the present gate is well calibrated, that abstention is cheap (false-veto rates of $0.52$ and $0.87$), that aggregation restores a correct decision, or that anything here measures commercial uplift (Table~\ref{tab:scorecard}).

Formal specifications are in Supplementary Material, Appendix A, robustness checks and ablations in Appendix D, the extended exposition in Appendix E, the proofs of Section~\ref{subsec:when-abstain} in Appendix F, and the extended text of each section in Appendix G, under the same title with its section number in parentheses.

\section{Related Work and the Gap}\label{sec:related-work}

This work sits at the intersection of seven literatures. \textbf{Structural demand estimation} \citep{berry1994estimating, berry1995automobile, nevo2001measuring} identifies substitution patterns and market power from aggregate market shares under strong assumptions about functional form and instrument availability; it targets the estimate itself and is agnostic about when a firm should act on it. \textbf{Vertical pass-through} \citep{villasboas2007vertical, nakamura2010accounting} documents that retailers rarely transmit a wholesale price change to the shelf one-for-one, which motivates treating pass-through as a process rather than a constant (Supplementary Material, Appendices A.7 and B). \textbf{Non-exchangeable conformal inference} \citep{gibbs2021adaptive, barber2023conformal} supplies the machinery used to bound the volume response under a chronological, non-stationary split (Appendix A.4). \textbf{Selective prediction and learning-to-defer} \citep{elyaniv2010foundations, mozannar2020consistent} formalize the principle that a system can decline to answer rather than answer badly, and the ACT/WAIT gate is an instance of it applied to a pricing decision; the contribution here is the instantiation, not the general principle. \textbf{Multiway panel clustering} \citep{chiang2022multiway} informs the resampling choices audited in Appendix A.1. And \textbf{performative prediction} \citep{perdomo2020performative}, alongside the critique that a relation estimated under one policy regime need not survive a new one \citep{lucas1976econometric, mackenzie2006engine}, names the risk that once this system moves prices the observational identification it relies on will erode, a governance problem this paper flags rather than solves (Appendix E).

A seventh, more practically oriented literature reports enterprise deployments of pricing and revenue-management systems \citep{talluri2006theory, marn1992managing, hormby2010marriott, deng2023alibaba, llenas2026pepsico}, typically a revenue or margin gain. It is also the sharpest illustration of the gap: none of these reports which of the system's identifying assumptions were checked, whether the reported gain would survive an audit of confounding, or at what level of aggregation the underlying estimates were trustworthy. The comparison is scoped to the three cited here and is not offered as a survey finding. Algorithm aversion \citep{dietvorst2015algorithm, dietvorst2018overcoming} suggests why that omission is costly: trust in an automated recommendation is fragile once the system has been seen to err, and is rebuilt by visible user control over an imperfect system, of which a declared and costed abstention is the clearest form.

\textbf{The gap.} The seven literatures above leave three gaps open. \emph{(i) Abstention as an object of design.} None of them treats the decision to abstain as a first-class object with its own diagnostics, its own measured cost, and its own conditions for reversal; the closest, selective prediction, formalizes the principle in a general classification setting and does not address what the relevant diagnostics are in an observational pricing panel with retail intermediation. \emph{(ii) Tacit assumptions as executable diagnostics.} None converts the assumptions a causal pricing system ordinarily leaves tacit, that controls are pre-treatment, that an instrument acts through one channel, that a comparison group is inert, that no adjacent optimizer touches the same unit, into diagnostics the system itself executes and reports. \emph{(iii) The legitimate unit of decision.} None specifies the conditions under which a decision illegitimate at one level of commercial aggregation becomes legitimate at a coarser one, or measures how far aggregation actually closes that gap. Sections~\ref{subsec:decision-layer}, \ref{subsec:tacit} and \ref{subsec:unit-of-identification} address the three gaps in that order, which are the three questions posed in Section~\ref{sec:intro}.

\section{Institutional Setting and Data}\label{subsec:data}

The valuation dataset is a weekly panel at the presentation~$\times$~region~$\times$~week level, integrating sell-in, market-audit sell-out, distributor point-of-sale, competitor pricing, promotional schedules and variable costs, with weather and holiday covariates. Gap imputation and publication rules belong to an upstream data contract, and the system's only diagnostic alert, for week-over-week volume drops above $80\%$ within a cell, is executable only against a production panel and is not exercised by any result in this paper.

The atomic unit is the \textbf{presentation} (brand $\times$ format family $\times$ pack size), not the SKU: list prices are assigned per presentation, so intra-presentation SKU variation carries no identifying price variance. The framework organizes around the structural hierarchy $\text{Category} > \text{Brand} > \text{Presentation}$ and assumes a portfolio architecture with multiple brands, distinct pack sizes, uninstrumented competitor prices and retail intermediation. The generators of Section~\ref{subsec:validation-design} instantiate one portfolio exhibiting that topology and the system is shown to run end to end on it; transfer to other portfolios sharing the structure remains a design expectation rather than a verified property. The \textbf{week} is the temporal unit of inference throughout, while the month governs only the decision cadence: that division is an identification decision and not a convenience, since under monthly aggregation the quota-pressure guard loses its contrast identically and the break tests fall from a hundred and twenty observations to twenty-eight.

\paragraph{Feasibility and admissibility.} A presentation~$\times$~region cell is \textbf{modelable} when five joint conditions hold on \emph{deflated} prices: a minimum coefficient of variation, at least four discrete price changes, bounded price--promotion and price--competitor correlations, and a non-frozen shelf status, with a minimum of forty usable observations. A presentation is then \textbf{admissible} when the diagnostic passes in at least half its regions, the treatment partial $R^2$ after residualization exceeds $0.10$, and the bootstrap interval width for the raw elasticity falls below $0.6$. The gate is evaluated together with the eight guards and not ahead of them (Definition~\ref{def:extended-classification}). Non-modelable units inherit a brand-level pooled elasticity and are marked non-actionable. (The five cutoffs are in Supplementary Material, Appendix E.)

\paragraph{A public real-data complement.} Every panel in this paper is synthetic except for two narrow checks (Section~\ref{subsec:real-data}) that run the estimation-and-interval core, and nothing else, on the public Dominick's Finer Foods scanner data \citep{kilts_dominicks}: weekly store-level prices and sales in a supermarket chain, not an intermediated channel. Neither has a known truth. The first tests the consistency of the system's intervals; the second compares its estimates with the responses to real price changes, an approximate truth that is not an experiment. Neither tests recovery.

\section{The Causal Governance Framework}\label{sec:methodology}

\begin{figure}[htbp]
\centering
\pandocbounded{\includegraphics[keepaspectratio,alt={The Pricing Intelligence system}]{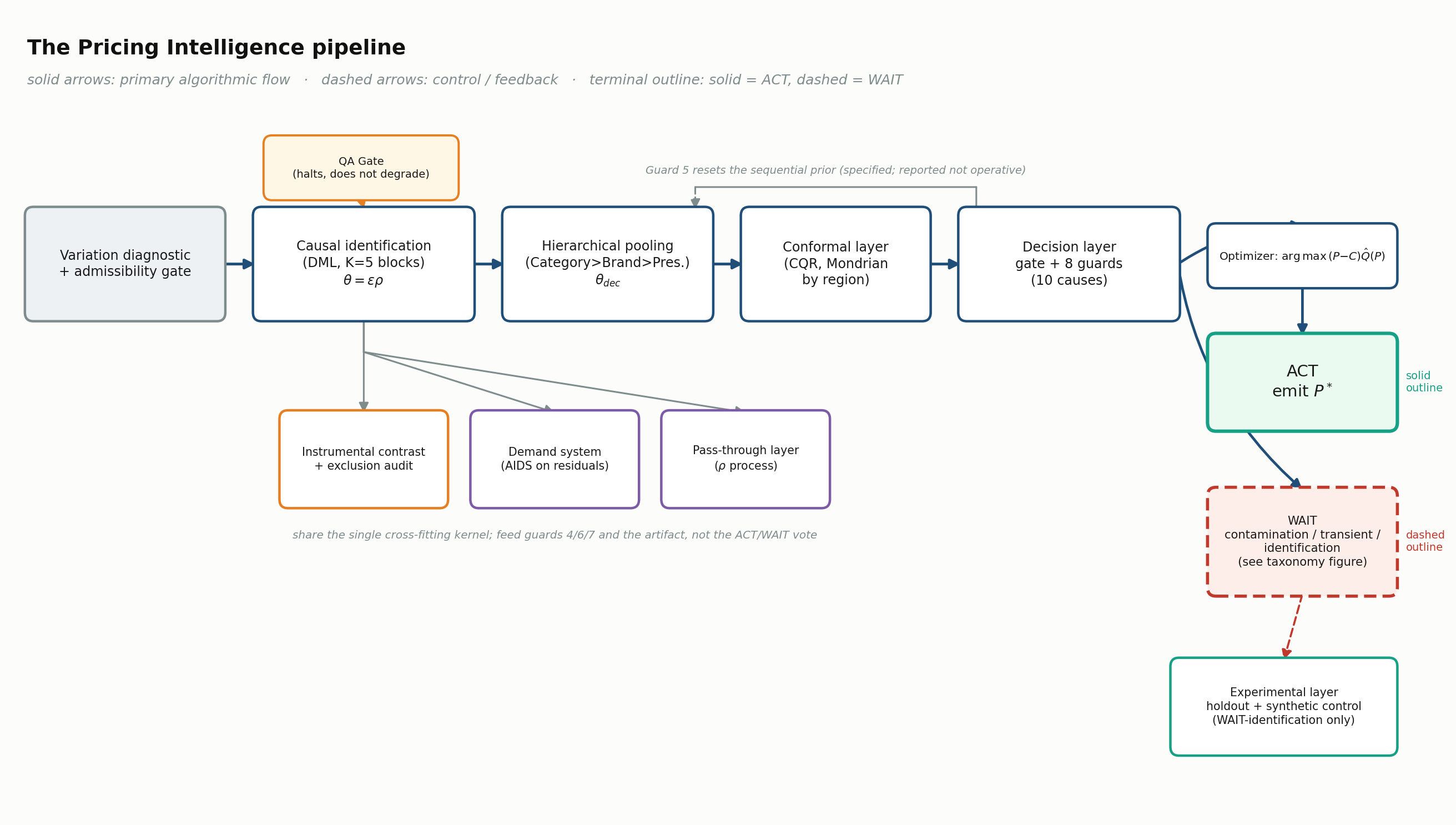}}
\caption{Schematic overview of the system. Solid arrows represent primary algorithmic flow,
dashed arrows denote control and feedback mechanisms, and the two terminal verdicts are
distinguished by outline: solid for ACT, dashed for WAIT.}\label{fig:pipeline}
\end{figure}

The system executes a sequential, artifact-driven process: diagnosis, causal identification, hierarchical pooling, conformal guarantees, and decision optimization (Figure~\ref{fig:pipeline}). Supplementary Material, Table G.1 states each layer in one line together with the verdict the validation run returned on it; formal specifications are in Appendix A, and the extended exposition from which this section is condensed is in Appendix E.

\subsection{The Estimation Layer: A High-Level Summary}

Identification rests on Double/Debiased Machine Learning: a partialling-out estimator with contiguous-block cross-fitting recovers the list-price elasticity $\theta = \varepsilon\rho$ --- the convolution of consumer response $\varepsilon$ and shopkeeper pass-through $\rho$ --- net of promotional, competitive, seasonal and calendar controls, each with a declared pre/post-treatment status enforced at estimation time. An instrumental contrast against cost shocks and a competitive-bracket decomposition probe what residualization on observables cannot reach; a hierarchical partial-pooling step shrinks each estimate toward its Category $>$ Brand $>$ Presentation branch but never itself unlocks a movement, since the ACT/WAIT gate votes solely on the raw estimate $\hat\theta$; and a conformal layer, Mondrian-stratified by region on a strictly chronological split, supplies the predictive interval on volume that the gate and guards consume. The $(S,s)$ repricing architecture is specified but not estimated at the region$\times$week grain available here, and a cash-pressure screening test stands in for the fuller two-part model that this panel's grain cannot support. Derivations, declared approximations and failure modes are in Supplementary Material, Appendices A to C.

\subsection{The Decision Layer}\label{subsec:decision-layer}

The decision layer adds no analysis. It converts what has been
estimated into one verdict per presentation and makes the optimizer act only where
identification holds. It is the least common component in enterprise pricing systems and the
one on which organizational trust turns \citep{dietvorst2015algorithm, dietvorst2018overcoming}. ACT requires passing every \emph{operative} guard simultaneously;
a guard that cannot be evaluated is reported as such rather than counted as passed, so a
presentation reaching ACT on seven evaluable guards is distinguishable from one reaching ACT on
eight. A verdict of WAIT, however, is not a single state: the decision layer partitions the
abstention causes into three mutually exclusive types, which makes the abstention diagnostic
rather than terminal, since each type states what would have to change for the presentation to
become actionable, and in one of the three cases that change is an experiment this system can
specify.

\begin{definition}[Three-way classification of guard-level abstention causes]\label{def:extended-classification}
The admissibility gate and the eight guards define ten guard-level abstention causes, since
Guard~1 evaluates two conditions separately: a regime shift in $\theta$ on excluding broken
weeks (\emph{1a}) and an excessive share of weeks so flagged (\emph{1b}). Let
$\mathcal{S}$ be the vector of the ten diagnostic outcomes for a presentation in a cycle, all of
them evaluated before any verdict is issued. The ten
causes partition into three disjoint subsets:
\begin{itemize}\tightlist
\item $\mathcal{G}_{\text{ID}}$ (\textsc{identification}, five causes): failing admissibility;
Guard~1a; Guard~1b; Guard~2; Guard~3. Each is a precision shortfall, or natural variation
contaminated by reactive pricing, that independently randomized perturbation directly resolves.
\item $\mathcal{G}_{\text{CONT}}$ (\textsc{contamination}, four causes): Guard~4; Guard~6;
Guard~7; Guard~8. Guard~4 belongs here rather than with precision because a strong-instrument
divergence is direct evidence of residual endogeneity in the observational variation itself,
by the same logic already applied to Guard~7. Each signals a specification or validity problem
that an experiment does not, by itself, repair.
\item $\mathcal{G}_{\text{TRANS}}$ (\textsc{transient}, one cause): Guard~5. Neither more
identifying variation nor a specification fix addresses a regime that has just ended; the
estimate becomes usable again once the new regime has accumulated its own history.
\end{itemize}
The decision layer is the mapping
$\mathcal{D}(\mathcal{S}) \rightarrow \{\text{ACT},\ \textsc{wait-contamination},\
\textsc{wait-transient},\ \textsc{wait-identification}\}$ defined by
\[
\mathcal{D}(\mathcal{S}) =
\begin{cases}
\text{ACT}, & \mathcal{G}_{\text{CONT}} = \emptyset \;\wedge\; \mathcal{G}_{\text{TRANS}} = \emptyset \;\wedge\; \mathcal{G}_{\text{ID}} = \emptyset,\\[4pt]
\textsc{wait-contamination}, & \mathcal{G}_{\text{CONT}} \neq \emptyset,\\[4pt]
\textsc{wait-transient}, & \mathcal{G}_{\text{CONT}} = \emptyset \;\wedge\; \mathcal{G}_{\text{TRANS}} \neq \emptyset,\\[4pt]
\textsc{wait-identification}, & \mathcal{G}_{\text{CONT}} = \emptyset \;\wedge\; \mathcal{G}_{\text{TRANS}} = \emptyset \;\wedge\; \mathcal{G}_{\text{ID}} \neq \emptyset.
\end{cases}
\]
\end{definition}

The classification is exhaustive but not symmetric across its three abstention outcomes.
\textsc{wait-contamination} and \textsc{wait-transient} both close the cycle without recourse to
more variation, since neither an experiment nor further observation repairs a validity failure
or a regime that has just changed. \textsc{wait-identification} is different in kind: it
certifies that it is specifically the available \emph{observational} variation that cannot
resolve the effect, which is exactly the condition under which independently assigned price
variation can, and it is used throughout this paper as formally equivalent to an \emph{Experiment
Candidate} (Table~\ref{tab:decision-taxonomy}; the ten causes grouped by subset are drawn in
Supplementary Material, Figure G.1).

Every cause in $\mathcal{S}$ is evaluated for every presentation before the decision is taken,
so a presentation that fails admissibility may also bind a contamination-type guard, and the
precedence then routes it to \textsc{wait-contamination} and not to an experiment: a single
contamination-type cause signals an assumption violation that independent randomization alone
does not repair, and a recent structural break makes the current estimate untrustworthy
independent of its precision. The definition adds no guard, no gate
and no threshold: it completes an assignment the framework already applied informally to
some causes, and it names a third category for the one cause, a
structural break, that fits neither of the first two.

{\small\begin{longtable}[]{@{}
  >{\raggedright\arraybackslash}p{(\linewidth - 6\tabcolsep) * \real{0.24}}
  >{\raggedright\arraybackslash}p{(\linewidth - 6\tabcolsep) * \real{0.30}}
  >{\raggedright\arraybackslash}p{(\linewidth - 6\tabcolsep) * \real{0.46}}@{}}
\caption{The four verdicts of the ACT/WAIT/EXPERIMENT decision layer under Definition~\ref{def:extended-classification}, in order of precedence. Every presentation in every cycle receives exactly one; only \textsc{wait-identification} hands off to a live experimental action rather than closing in a terminal block.}\label{tab:decision-taxonomy}\\
\toprule\noalign{}
Verdict & Nature of the block & What would change it \\
\midrule\noalign{}
\endfirsthead
\multicolumn{3}{@{}l}{\itshape Table \thetable{} (continued)}\\
\toprule\noalign{}
Verdict & Nature of the block & What would change it \\
\midrule\noalign{}
\endhead
\bottomrule\noalign{}
\endlastfoot
ACT & None & Identification and stability both hold; the optimizer emits $P^*$ \\
\textsc{wait-contamination} & Structural: a validity assumption fails & Control of the offending channel, not more variation; an experiment alone does not repair it \\
\textsc{wait-transient} & Temporal: the regime just changed & Accumulated history under the new regime; the estimate becomes usable again without any intervention \\
\textsc{wait-identification} \emph{(Experiment Candidate)} & Observational: the available variation cannot resolve the effect & Independently assigned price variation. This is the non-terminal verdict that hands the presentation off as a formal candidate for portfolio experimental design \\
\end{longtable}}

The admissibility gate and the eight guards that populate this taxonomy are stated in Supplementary Material, Table E.1 and developed in Appendix E. Guards 1--5 address the quality and currency of the estimate (contaminated weeks, the Cinelli--Hazlett robustness value, the stability of the recommendation across the width of the uncertainty, and structural-break tests); Guards 6--8, added after an external methodological audit, address the ways an estimate can be well measured and still measure something other than what it claims, including a quota-pressure bracket that treats a neighboring commercial order-recommendation system as a possible confounder. Section~\ref{subsec:tacit} states, for each of these diagnostics, whether it adapts an established econometric tool or has no close equivalent in that literature. Guard~5 is computed and classified on every cycle but is not an active veto in the validated configuration, a retirement documented in Section~\ref{subsec:limitations}.

\subsection{When does withholding beat acting?}\label{subsec:when-abstain}

Whether the decision layer abstains is one question; whether abstaining is the right decision is another, and it has a short answer. This subsection gives it for the choice between acting on the optimizer's recommendation and holding the current price, the two options whose outcome is a contribution. What a \textsc{wait-identification} verdict is worth as an experiment is outside it. The statements are decision theory, not econometrics: they take the estimator as given, in the tradition of statistical treatment rules \citep{manski2004statistical}. Proofs are in Supplementary Material, Appendix~F.

\paragraph{Setting.} Take one presentation at one decision date. Let $x_0$ be the log of the current price, $\mathcal{P}$ the corridor in log price, $C$ the true expected contribution as a function of log price, which the system does not know, and $\hat x\in\mathcal{P}$ the price the optimizer would emit if the system acts. Acting yields the gain $\Gamma=C(\hat x)-C(x_0)$ over holding, and \textsc{wait} yields zero. Let $\mathcal{S}$ be what the system observes at the decision date, the estimates and the diagnostic outcomes of Definition~\ref{def:extended-classification}, and let $A\in\{0,1\}$ be the indicator of ACT, a function of $\mathcal{S}$. With $x^\ast$ the maximizer of $C$ on $\mathcal{P}$, the expected decision loss against the best price in the corridor is $\mathbb{E}[C(x^\ast)-C(x_0)]-\mathbb{E}[A\,\Gamma]$.

\begin{proposition}[Value of an abstention rule]\label{prop:value}
Let $q=\Pr(A=1)\in(0,1)$. (i) The rule has a lower expected loss than never acting if and only if $\mathbb{E}[\Gamma\mid A=1]>0$, and a lower expected loss than always acting if and only if $\mathbb{E}[\Gamma\mid A=0]<0$. (ii) Its expected loss equals that of a rule that acts at random with the same probability $q$, independently of $\Gamma$, minus $\mathrm{Cov}(A,\Gamma)=q(1-q)\{\mathbb{E}[\Gamma\mid A=1]-\mathbb{E}[\Gamma\mid A=0]\}$. (iii) Among rules that are functions of $\mathcal{S}$, $A^\ast=\mathbf{1}\{\mathbb{E}[\Gamma\mid\mathcal{S}]>0\}$ minimizes the expected loss. The identity in (ii) also holds for the probability of harm, $\Pr(A=1,\Gamma<0)$, with $\mathbf{1}\{\Gamma<0\}$ in place of $\Gamma$.
\end{proposition}

Proposition~\ref{prop:value} assumes nothing about the estimator or the demand. It says that an abstention rule earns something over acting at random only through the covariance between its decisions to act and the gains from acting: it has to keep the presentations where acting pays and drop those where it does not. A rule whose abstentions fall where acting pays on average loses to always acting.

\begin{proposition}[Headroom against error]\label{prop:headroom}
Suppose that on the corridor $C(x)=C(x^\circ)-\tfrac{\kappa}{2}(x-x^\circ)^2$ with $\kappa>0$, where $x^\circ$ is the unconstrained maximizer, possibly outside $\mathcal{P}$. Write $d=x^\circ-x_0$ for the headroom, $\hat m=\hat x-x_0$ for the recommended move and $e=\hat x-x^\circ=\hat m-d$ for the error of the recommended price. Then $\Gamma=\tfrac{\kappa}{2}(d^2-e^2)=\tfrac{\kappa}{2}\hat m(2d-\hat m)$, and for any set $\mathcal{G}$ of presentations determined by $\mathcal{S}$ with $\mathbb{E}[\kappa\hat m^2\mid\mathcal{G}]>0$,
\[
\mathbb{E}[\Gamma\mid\mathcal{G}]=\tfrac12\,\mathbb{E}[\kappa(d^2-e^2)\mid\mathcal{G}]=\tfrac12\,\mathbb{E}[\kappa\hat m^2\mid\mathcal{G}]\,(2\beta_{\mathcal{G}}-1),\qquad
\beta_{\mathcal{G}}=\frac{\mathbb{E}[\kappa\hat m\,d\mid\mathcal{G}]}{\mathbb{E}[\kappa\hat m^2\mid\mathcal{G}]}.
\]
Acting therefore beats holding on $\mathcal{G}$ if and only if the mean-square headroom exceeds the mean-square error of the recommended price, $\mathbb{E}[\kappa d^2\mid\mathcal{G}]>\mathbb{E}[\kappa e^2\mid\mathcal{G}]$, equivalently if and only if the calibration slope $\beta_{\mathcal{G}}$ of the true headroom on the recommended move exceeds $1/2$.
\end{proposition}

Proposition~\ref{prop:headroom} puts the condition in terms the decision layer can be judged by. A recommendation can be noisy, biased or attenuated and still be worth acting on, provided the true headroom regressed on the recommended move has a slope above one half; one that rests on a perfectly identified estimate loses when its error exceeds the headroom it is meant to capture. Identification matters only through the slope. The slope is that of the Mincer--Zarnowitz calibration regression \citep{mincer1969evaluation}, and the threshold of one half is the elementary condition under which a forecast beats a null forecast under squared loss, the criterion behind out-of-sample $R^2$ \citep{campbell2008predicting}; what the proposition adds is its reading as a rule for a pricing decision, where the null is holding the price and the loss is the weighted square $\kappa(d^2-e^2)/2$, and its check against the records of a decision system (Section~\ref{subsec:condition-check}).

\begin{corollary}[Classes]\label{cor:classes}
Partition the presentations into classes by a function of $\mathcal{S}$, such as identifiable or not, and let $g_c=\mathbb{E}[\Gamma\mid c]$. For a rule that, given the class, acts with probability $q_c$ independently of $\Gamma$, $\mathrm{Cov}(A,\Gamma)=\sum_c\Pr(c)(q_c-q)\,g_c$. If $g_c>0$ in every class, which by Proposition~\ref{prop:headroom} is the case if $\beta_c>1/2$ in every class, no such rule beats always acting.
\end{corollary}

\begin{proposition}[Direction]\label{prop:direction}
Suppose $C$ is strictly increasing on $\mathcal{P}$ up to its maximizer $x^\ast$ and strictly decreasing after it, and $x^\ast>x_0$. Then $\Gamma<0$ for every $\hat x<x_0$ and $\Gamma>0$ for every $x_0<\hat x\le x^\ast$; if $x^\ast$ is the upper bound of the corridor, $\Gamma>0$ for every $\hat x>x_0$. The same holds with the directions reversed if $x^\ast<x_0$, and if $x^\ast=x_0$ acting never gains.
\end{proposition}

When the corridor binds, then, only the direction of the recommended move decides the sign of the gain, not its size, which is why acting is almost always right in a market whose optimum is the corridor's ceiling (Section~\ref{subsec:baselines}).

The quadratic form in Proposition~\ref{prop:headroom} is a local model, adopted because it makes the condition exact, not a claim that demand is quadratic; Section~\ref{subsec:condition-check} checks the condition on generators whose true curve is not. The slope $\beta$ cannot be observed outside a simulation, since it needs the true optimum. That is why the experimental arm of the decision space matters here: a designed price experiment is the route by which the calibration of recommended moves could be learned in practice.

\subsection{System Architecture}

Five principles are enforced in the implementation rather than recommended in documentation: a single residualization kernel feeding the causal estimator, the demand system and the pass-through deconvolution; a single temporal resampling policy; labeled rather than disguised placeholders; no undeclared control; and one effective unit of inference, so that any statistic whose distribution depends on a sample size counts independent periods rather than panel rows. Every cycle's output is a versioned, schema-validated artifact whose verdict field is closed over exactly the four states of Definition~\ref{def:extended-classification} (Supplementary Material, Appendix E).

\subsection{Validation Design}\label{subsec:validation-design}

Every quantity reported below, with one exception, is measured against a data-generating process whose answers are known by construction; the exception is Section~\ref{subsec:real-data}, which runs the estimation-and-interval core on public scanner data with no known truth. The premise is that a methodological claim about identification is worth what its recovery experiment shows: a system that abstains correctly on a panel where the truth is unknown is indistinguishable from one that abstains at random. The \textbf{five-regime generator} simulates $W=120$ weekly periods across six regions and six presentations spanning three brands and two categories, with five identification regimes that vary the amount and kind of price variation available, from a \emph{clean} regime with ample list movement to regimes built deliberately to fail on weak variation, confounding with promotion, or collinearity with the competitor. A generator with known answers is also one with chosen answers, so the run uses a second, the \textbf{extended generator}, with eight presentations and wider variation in pass-through, regime assignment and elasticity magnitude, and reports both wherever they disagree; a third generator, written independently of both, is used only in Section~\ref{subsec:other-market}. Validation runs in four layers (statistical, commercial sanity, back-testing and stress) plus an experimental layer (a deterministic holdout, a synthetic control \citep{abadie2010synthetic}, permutation inference, and a contamination test of the no-interference half of SUTVA \citep{rubin1980randomization}); every parameter and the protocol are stated in Supplementary Material, Appendix E.

\paragraph{Bookkeeping.} Three conventions govern how every figure below should be read. \emph{Pre-registered verdicts}: thirty decision rules were written before the run, each pairing a claim the manuscript makes with a threshold that would settle it; ten pass, nineteen fail, and one does not apply because the failure it was written to adjudicate did not reproduce. Five of the thirty were rewritten between the first and second drafting of the register because they were mis-specified, not because their numbers were unwelcome; the correction is to the rule and not to the measurement (Supplementary Material, Table D.2 lists all thirty against their declared thresholds). These rules are engineering verification, internal to this run and distinct from the paper's three contributions (Section~\ref{sec:intro}). Most of the nineteen failures are not evidence of malfunction, since proving that an estimator cannot recover the injected effect under collinearity, a weak instrument, or a confounded design is itself an identification result; one failure is not of that kind: V9 measures what abstention costs on presentations that deserve an answer, and its failure is a cost of the system as configured, reported in Section~\ref{subsec:funnel}. \emph{Separated seeds}: where a design choice was selected by measurement, the selection ran on one set of seeds and the published figure on another. \emph{No single $R$}: the replication count is set per experiment, from $300$ seeds down to $5$ (Table D.1).

\section{Audit of Seven Tacit Assumptions}\label{subsec:tacit}

The strongest claim this work supports is not the integration of causal econometrics into a
pricing workflow, which is by now a well-populated genre \citep{chernozhukov2018double}, but a
pattern that emerged from auditing a carefully built system and then executing it. The failure
modes that survive a competent build are rarely errors of estimation. They are unstated
assumptions about the \emph{environment}, and each can be converted into a quantity that costs
less to measure than its consequence costs to absorb. Table~\ref{tab:tacit} states the seven
found here, four by reading the design and three by running the validation, and is the second
contribution in compact form: the columns give what a decision-maker would otherwise take
for granted, the contrast that tests it, and what would have gone wrong silently had the test not
been run.

\begin{table}[htbp]
\centering\small
\caption{Seven assumptions that a careful build leaves tacit, and the executable contrast each
becomes. The upper block was found by auditing the design and is executed by the system; the lower
block was found by executing the validation and is audited against a known truth before
deployment, its outcomes governing how the system's outputs may be published. In every row the test is cheaper than the consequence of leaving the assumption
implicit, which is the property that makes the pattern worth generalizing past
pricing.}\label{tab:tacit}
\begin{tabular}{@{}p{0.26\linewidth}p{0.32\linewidth}p{0.34\linewidth}@{}}
\toprule
Tacit assumption & Executable test & Consequence if left implicit \\
\midrule
\multicolumn{3}{@{}l}{\emph{Found by auditing the design: executed by the system}}\\[2pt]
The controls are pre-treatment
  & Declared temporal status per column; estimation aborts on an undeclared one; mediator bracket (Guard 6)
  & Conditioning on a mediator inflates $|\theta|$; the engine under-recommends increases and the number stays plausible \\
The instrument acts through one channel
  & Exclusion audit: residual correlation of instrument and competitor price against a set carrying an explicit trend (Guard 7)
  & A common cost shock moves both estimators together, so the Hausman-type contrast is small \emph{because} the restriction fails \\
The comparison group is inert
  & Contamination test on the control group's price shift, against a permutation distribution of false cuts and an economic floor
  & A competitor reading regional prices contaminates the donor pool; the measured effect is attributed to the intervention \\
No other optimizer acts on the same unit
  & Quota-pressure bracket with a declared source hierarchy (Guard 8)
  & End-of-period effort lowers effective price and raises volume together; the engine learns its own firm's commercial policy as demand \\[4pt]
\multicolumn{3}{@{}l}{\emph{Found by executing the validation: audited against a known truth before deployment}}\\[2pt]
A computed interval is an honest interval
  & Coverage against a known truth; conditional-versus-unconditional dispersion across designs
  & The bootstrap targets dispersion conditional on the realized design, which is not what a guard certifies; three governance switches appeared to operate and did not \\
The unit of decision is the unit of identification
  & Aggregate the estimate to brand and category and measure the dispersion across replications
  & Design bias averages at the square-root rate, so the RMSE falls from $0.488$ at the presentation level to $0.159$ at the category level under the sieve learner, and from $0.571$ to $0.336$ under the boosted one; publishing per presentation publishes at a grain the evidence does not support \\
A threshold calibrated on one configuration is a threshold
  & Freeze the cutoffs and sweep the data-generating configuration
  & Gate precision (the share of admitted cells with a usable estimate) is $0.5$ on the calibration configuration and between $0.000$ and $0.75$ outside it (undefined where the gate admits nothing), against a rule of $\ge 0.80$ \\
\bottomrule
\end{tabular}
\end{table}

The symmetry of the table is the argument. Each row names an assumption that nothing in the
estimation stage can detect, because each is a statement about the environment rather than about
the data; each is answered by an executable contrast, computed by the system in the first four rows and
audited against a known truth in the last three; and in each case the
test is cheap while the consequence is not. Because they need a known truth, the last three are not executed on a production panel; what carries over to deployment is what they decided: what interval a guard may certify, at which unit to publish, and that thresholds are published as a surface of precision and recalibrated on a real category. The sixth row also stands against this paper's
own account: the same work argues for estimating at the week rather than the
month, because the grain of a panel is an identification question, and then publishes per
presentation without asking the same question of the cross-section.

Provenance differs across the seven. Four reuse established tools: the exclusion audit (Guard~7) turns instrumental-variable exogeneity \citep{rossi2014even} into a residual-correlation contrast; the contamination test adapts permutation inference on the no-interference half of SUTVA \citep{rubin1980randomization}; the threshold check adapts out-of-configuration validation; and the coverage diagnostic reuses conformal machinery \citep{gibbs2021adaptive, barber2023conformal}. Two have no close equivalent in the literatures this paper compares itself against: the quota-pressure bracket (Guard~8), since interference between independently operated optimizers acting on one unit is not a standard identification concern, and the aggregation test of row six (Section~\ref{subsec:unit-of-identification}). The mediator bracket (Guard~6) sits between: the bad-control concern is well known, and what is new is its enforcement by a declared temporal status per column.

\section{Legitimate Commercial Aggregation Rule and Key Results}\label{sec:results}

This section reports what the validation run measured; a layer is reported as validated only where an experiment measured its claim against a known truth, and Supplementary Material, Table G.2 gives the verdict the run returned for each layer. Sections~\ref{subsec:coverage-experiment} to~\ref{subsec:unit-of-identification} carry the third contribution, Sections~\ref{subsec:impossibility} to~\ref{subsec:extended-taxonomy} and~\ref{subsec:baselines} to~\ref{subsec:condition-check} the first, Section~\ref{subsec:real-data} the checks on real data, and Section~\ref{subsec:recovery} the recovery experiments on which the rest builds. The ablation battery behind these results, eleven contrasts against alternative nuisance learners, cross-fitting schemes, point estimators, sign filters, and outcome variables, is reported in Supplementary Material, Appendix D.

Table~\ref{tab:scorecard} summarizes, before the detail, which claims of the paper the validation supports and which it does not.

\begin{table}[htbp]
\centering\footnotesize
\caption{What the validation supports and what it does not. \emph{Synthetic}: measured against a known truth; \emph{public}: Dominick's data. The condition and the propositions were derived after the experiments of Sections~\ref{subsec:baselines} and~\ref{subsec:other-market}.}\label{tab:scorecard}
\begin{tabular}{@{}p{0.30\linewidth}p{0.41\linewidth}p{0.21\linewidth}@{}}
\toprule
Claim & Evidence & Status \\
\midrule
A condition predicts when acting pays (calibration slope above $1/2$) & Sign of the gain predicted in $37$ of $38$ groups; rank correlation $0.92$ (Section~\ref{subsec:condition-check}) & Supported; synthetic; derived post hoc \\
Identifiability does not separate the groups where acting pays & In the main independent scenario no rule built on the identifiable classes beats always acting (Corollary~\ref{cor:classes}) & Supported; synthetic \\
Abstention is aimed where the condition fails & ACT rate $0.000$ in every group below one half (Table~\ref{tab:decomp}) & Supported; synthetic \\
Abstention is cheap enough & False-veto rate $0.52$ and $0.87$ against a registered limit of $0.20$ (Sections~\ref{subsec:funnel} and~\ref{subsec:other-market}) & Not supported \\
Abstention reduces harm & $0.099\to0.002$ and $0.357\to0.036$; a random rule at the same ACT rate gives $0.016$ and $0.050$ (Table F.2) & Supported; in the independent market mostly by acting rarely \\
Aggregation improves the point estimate & RMSE $0.488\to0.159$; square-root law within $7.4\%$ (Section~\ref{subsec:unit-of-identification}) & Supported; synthetic \\
Aggregation restores a correct decision & Decision function evaluated at the presentation level only; no interval both covers and clears the width gate (Section~\ref{subsec:impossibility}) & Not tested \\
Resampled intervals undercover & Replicated on Dominick's ($0.565$ and $0.762$ of pairs within their own errors) & Supported; synthetic and public \\
A single-window UPC estimate forecasts real price changes & Slope $0.22$, skill $-0.25$ (Section~\ref{subsec:real-data}) & Not supported; public \\
\bottomrule
\end{tabular}
\end{table}

\subsection{Recovery of the injected elasticities}\label{subsec:recovery}

Over $63$ replications of the five-regime generator (Supplementary Material, Table D.3), the \emph{clean} regime reaches admissibility in $0.968$ of replications and ACT in $0.944$, with an estimate of $-0.962$ against a truth of $-1.105$ (a bias of $0.143$ toward zero, at a coverage of $0.484$ against a nominal $0.95$), while every other regime reaches both in exactly $0.00$ and the estimator does not degrade gracefully: in \emph{collinear} the median estimate is $+0.248$, the wrong sign, and in \emph{confounded} and \emph{weak} it is $-0.103$ and $-0.183$ against truths of $-1.190$ and $-0.850$. On the extended generator (Table D.4; $53$ replications) the round-point and the confounded presentation are each admitted in $0.981$ of replications under the boosted learner, against $0.00$ on the five-regime generator, while the clean-regime bias transfers ($0.143$ against $0.197$). A dedicated experiment finds no large spurious estimate under a round-point shelf price ($-0.069$ against a truth of $0$). The competitive bracket cannot be checked on the panel as generated, where the injected cross term lies below the noise floor (signal-to-noise ratio $0.215$); on a variant in which the competitor follows the own price, the reaction function is recovered almost exactly and the cross elasticity at $49\%$ of its injected value (Supplementary Material, Appendix D).

\subsection{Why a resampled interval is not enough on its own}\label{subsec:coverage-experiment}

The block-bootstrap interval that the causal layer reports \citep{kunsch1989jackknife} undercovers against known truth, and none of seven alternative resampling and clustering schemes, at two nuisance learners, closes the gap: the best, multiway clustering, reaches $0.68$ under the boosted learner and $0.80$ under the sieve against a nominal $0.95$ (Table D.8). The shortfall is one of centering rather than width: coverage conditional on a realized design is $0.881$ for the boosted learner and $0.475$ for the sieve although the bootstrap is wide relative to the dispersion that remains within a design ($0.51$ and $0.41$ of its own standard error; four independently drawn designs with forty noise draws each, and relative to total dispersion the ratio is $0.62$ for the boosted learner and $2.04$ for the sieve, so the two learners are read separately). The conformal band does not share the problem: on $100$ replications at ten four-week rolling origins its mean out-of-time coverage is $0.909$ against a nominal $0.90$, falling from $0.95$ at the full panel to $0.79$ at the shortest history tested, $84$ weeks. An interval computed from a single realized panel therefore cannot always be certified as honest.

\subsection{What a single realized panel cannot tell you}\label{subsec:hierarchical-coverage}

Resampling weeks inside one realized panel cannot see the dispersion that would arise from having observed a different set of price movements altogether, what Section~\ref{subsec:unit-of-identification} treats as \textbf{design variance}, closely related to the design-based uncertainty of \citet{abadie2020sampling}. A hierarchical variance component estimated across presentations, each effectively a different realized design \citep{paule1982consensus}, can: on a simulated replica of this panel (Table D.9; $70$ replications on eight presentations) the posterior-predictive construction reaches coverage of $0.914$ when every presentation shares the same truth and $0.92$ when truths differ, against the bootstrap's $0.484$ in both, which is direct evidence that the missing variance is design variance; the posterior point estimate alone covers \emph{worse} than the bootstrap ($0.407$ and $0.42$). None of this is adopted in the production configuration reported throughout this paper, a non-adoption declared as a limitation in Section~\ref{subsec:limitations}.

\subsection{The unit at which this channel identifies}\label{subsec:unit-of-identification}

If part of what defeats a per-presentation interval is design variance, it should average out
across units whose designs are independent, at a rate the arithmetic predicts, and this is the
paper's third contribution: showing when the unit of decision should move from the presentation
to the brand or the category.

Supplementary Material, Table D.5 aggregates from the presentation to the brand and to the category
over 70 replications, on eight presentations of one category that form two brands of four
(list-price elasticities $-1.105$ and $-0.85$). Under the boosted
learner the standard deviation of the estimate across
replications falls by a factor of $2.03$ at the brand, where four presentations are pooled, and
by $2.94$ at the category, where eight are; the factors predicted by independent design biases
averaging at the square root of the unit count are $2.00$ and $2.83$. Under the sieve learner
the measured factors are $2.02$ and $3.04$ against the same predictions, and the largest
relative deviation from the square-root law across the four contrasts is $7.4\%$. The mechanism
is therefore confirmed rather than argued.

Two consequences follow. The first is a usable
estimate: like for like, the root mean squared error falls
from $0.488$ at the presentation level to $0.159$ at the category level under the sieve learner ($-67\%$) and from $0.571$
to $0.336$ under the boosted learner ($-41\%$). The second is a boundary on where this
purchases anything: widening the panel in \emph{regions} does not average design bias, because
the list price is common across regions of the same presentation and the design is therefore
shared, whereas widening it in \emph{presentations} does.

Aggregation is not a free repair for the rest of what a decision needs. For the block-bootstrap
interval, coverage
does not improve with the level, at $0.487$, $0.400$ and $0.314$ across the presentation, brand and
category levels under the boosted learner and at $0.271$, $0.264$ and $0.257$ under the sieve, so
aggregation improves the point estimate without making the bootstrap interval honest, and whether
an interval at the brand or category level reaches nominal coverage inside a usable width is the
question of Section~\ref{subsec:impossibility}. The rule that follows is this: when
presentation-level evidence does not support a decision, publish the estimate at the brand level
(four presentations pooled in this panel) or the category level (eight), aggregating over
presentations and not over regions, because only presentations carry independent price designs; the
measured benefit is to the point estimate, and the rule does not by itself certify an interval. Aggregation therefore restores the accuracy of the estimate and, on Dominick's data, admissibility in $7$ of $11$ and $9$ of $11$ categories under the hierarchical interval (Section~\ref{subsec:real-data}); it is not shown to restore a correct decision, since the decision function is evaluated at the presentation level only, and on real data the averaging is weaker than the square-root law (the between-window dispersion of a UPC relative to that of its category-pooled estimate is $0.90$ against $2.53$ predicted; Supplementary Material, Appendix D).

\subsection{The limit on per-presentation precision in this panel}\label{subsec:impossibility}

In none of the twelve level-learner-scenario combinations tested (Supplementary Material, Table D.15) does an interval with coverage of at least $0.90$ fit inside the admissibility width threshold of Section~\ref{subsec:data}: the hierarchical predictive construction reaches that coverage in ten of the twelve, at between $1.55$ and $3.38$ times the threshold, and the block bootstrap fits inside the threshold in ten of the twelve but covers between $0.224$ and $0.455$. That is a property of how little independent price variation a hundred-and-twenty-week panel of this shape supplies, not a defect specific to this implementation, and it is the operational justification for treating WAIT as a first-class output: on some presentations, at some levels of aggregation, the evidence-honest answer is that the data does not yet support a per-item decision at the confidence a committee should require. Where the limit sits and what would close it is developed in companion methodological work \citep{delgado2026acrossdesignuncertaintyshortpricing}; Section~\ref{subsec:artifact} discusses what a committee should do while it stands.

\subsection{Detectable effect, and how far the gate travels}\label{subsec:mde}

Across six candidate learners, the injected magnitude at which the detection rate (the block-bootstrap interval excluding zero) first crosses $80\%$ from below is $0.9243$ under the production-adjacent boosted-tree learner, close to the top of the range swept ($0.1$ to $1.0$ at $W=120$), while the lasso, ridge and the sieve reach it at $0.5$ or below; the textbook frontier is optimistic relative to the honestly simulated one by a factor of $2.335$ for the boosted learner, against $1.057$ and $1.045$ for ridge and the sieve (the figure and the full sweep: Supplementary Material, Appendix D).

\subsection{Does the gate generalize beyond the configuration it was calibrated on?}\label{subsec:gate-generalization}

Freezing the thresholds at the values calibrated on configuration A and moving the data-generating configuration (Supplementary Material, Table D.10; variants B, C and D) gives, under the boosted learner, a gate precision of $0.5$ in A, undefined in B because the gate admits nothing at all, $0.0$ in C and $0.75$ in D, and under the sieve learner $0.5$, $0.0$, $0.5$ and $0.6$: outside A the worst precision, wherever the gate admits anything at all, is exactly $0.0$, against a rule of $0.80$. The near-perfect separation on A is therefore an artifact of calibration and does not travel as a general claim; what travels is the surface of precision over the configuration space, to be published and recalibrated on a real category, and the gate is to be described as calibrated and not as solved.

\subsection{The abstention funnel}\label{subsec:funnel}

The funnel over 50 replications and $300$ presentation evaluations (Supplementary Material, Table D.11) applies the admissibility gate and Guards 1 through 4; Guards 5 through 8 are applied downstream, in the full guard pipeline. It ends at an ACT rate of $0.307$, nearly equal to the $0.333$ of the portfolio that is identifiable by construction, a statement about those four guards and not about the system: the taxonomy run of Section~\ref{subsec:extended-taxonomy}, with all ten causes in force, sees the full guard set remove a further share of the admitted presentations ($17.6\%$ reaching ACT against $32.4\%$ admitted).

\paragraph{What the guards cost on a panel that deserves an answer.} A funnel on a portfolio that is two-thirds non-identifiable cannot distinguish discipline from paralysis, so the run adds the measurement on the decision function the system actually runs, on presentations identifiable by construction, so that every veto is a false positive: $400$ evaluations in $200$ replications of the full pipeline (Supplementary Material, Table G.3). The admissibility gate with Guards 1 to 4 vetoes $58$ of the $400$, a false-veto rate of $0.145$ ($95\%$ interval $0.113$ to $0.182$), inside the pre-registered limit of $0.20$. The full decision function, which applies Guards 6, 7 and 8 as well (Guard~5 is recorded and not operative), vetoes $209$ of the $400$, a rate of $0.522$ ($0.475$ to $0.570$): rule V9 fails when it is measured on the function the system runs. The cost is concentrated in two guards, the exclusion audit (Guard~7), which fires on $0.33$ of the identifiable presentations, and the mediator bracket (Guard~6), on $0.20$; removing both brings the rate to $0.145$. The system never acts on a non-identifiable presentation in this generator ($0$ of $800$ evaluations; Wilson upper bound $0.005$), which Section~\ref{subsec:other-market} shows does not carry over to a differently written market. Section~\ref{subsec:baselines} measures what this abstention costs in decisions.

\subsection{Does the taxonomy close the classification gap?}\label{subsec:extended-taxonomy}

Over $18$ independent realizations of the decision mechanism on panels disjoint from the funnel's draws ($108$ decision instances), $89$ return a WAIT of some kind, an $82.4\%$ block rate against an ACT rate of $17.6\%$. The informal five-cause assignment the framework carried before this paper resolves $96.6\%$ of those $89$; Definition~\ref{def:extended-classification} resolves $100\%$, closing the $3$ instances ($3.4\%$) whose only binding cause was one the earlier rule never named, Guard~4, without altering a single guard's threshold or binding rate. Contamination-type causes dominate the blocked set, $97.8\%$ against $2.2\%$ identification-type ($2$ of $89$ instances, the only ones eligible for the experimental route); the share reflects this population's guard incidence together with the precedence of contamination over identification, since $71$ of the $89$ blocked instances bind causes of both types, so the eligibility the taxonomy produces varies with the portfolio and is reported here on one population (per-cause incidence: Supplementary Material, Table D.20).

\subsection{Do the diagnoses survive real scanner data?}\label{subsec:real-data}

This subsection asks whether the two findings behind the abstention rule, that a resampled interval is too narrow and that the width gate leans on it, also appear on a public panel with no known truth: the Dominick's Finer Foods scanner data \citep{kilts_dominicks}, weekly store-level sales and shelf prices for eleven product categories. Both tests run only the estimation-and-interval core; neither tests recovery of a known parameter, and a supermarket chain is not the intermediated channel of Section~\ref{subsec:data}.

\paragraph{Consistency.} For each UPC (the counterpart of a presentation) and each window of $120$ consecutive observed non-promotion weeks, the slope of $\log(1+\text{units})$ on the log shelf price is estimated under both nuisance learners, with a block-bootstrap interval and the hierarchical interval of Section~\ref{subsec:hierarchical-coverage}. Two estimates of the same UPC from consecutive, disjoint windows should agree within their stated errors: $z=(\hat\theta_2-\hat\theta_1)/\sqrt{\hat s_1^2+\hat s_2^2}$ should have unit standard deviation, and $95\%$ of pairs should fall within $\pm1.96$. They do not. Across the nine categories with enough pairs ($1{,}181$ in all), the median standard deviation of $z$ is $2.50$ (sieve) and $1.64$ (boosted), and the median share of pairs within the bootstrap's own errors is $0.565$ and $0.762$; a between-window variance estimated out of sample across UPCs brings it to $0.934$ and $0.939$, though not in every category (Supplementary Material, Tables D.16 to D.19; per-category results: Table G.4). Under the width gate of $0.6$ the bootstrap would release $222$ of $1{,}908$ UPCs (sieve) and the hierarchical interval $49$, and the gate clears only at aggregated levels ($7$ of $11$ and $9$ of $11$ categories under the hierarchical interval), as the aggregation rule predicts. The test establishes that the bootstrap's undercoverage is not an artifact of the generators, not that the hierarchical interval is right: the added variance term is fitted to the same disagreement between windows, which mixes design variance with real change in the elasticity, so shares of $0.93$ to $0.94$ are a calibration check and not independent validation.

\paragraph{Do the estimates forecast real price changes?} The second test gives the estimates an approximate truth from the same files. An \emph{event} is a change of at least $3\%$ in one store's regular price, with four stable weeks either side, and its response is the change in log volume net of the stores that kept the price. The estimate of window $k$, $\hat\theta_k$, forecasts the response to an event in window $k+1$ as $\hat\theta_k\,\Delta\log P$, so the test is out of sample and forward in time; it is not an experiment, since the price changes are the retailer's, and the event definition, gate thresholds and five contrasts were fixed in the code before its first full run. Over $10{,}676$ events in $519$ UPCs, the regression of the realized response on the forecast has a slope of $0.22$ ($1$ would be calibration), and the forecast skill, an out-of-sample $R^2$ against a null forecast \citep{campbell2008predicting}, is $-0.25$ ($-0.34$ to $-0.17$) against predicting no response and $-0.20$ ($-0.28$ to $-0.13$) against the median estimate of the same category. The width gate passes $21$ of the UPCs with events, and the gate that adds the between-window variance passes none, so four of the five contrasts fixed beforehand cannot be evaluated and the fifth runs contrary to what the rule predicts; the decision metric gives the estimates no support either (Supplementary Material, Tables D.35 to D.39). Taken at face value, a window's UPC-level estimate is not a usable forecast of the response to a real price change in this chain. The slope of $0.22$ (upper limit $0.29$) is far from calibration and below the one half at which, in Proposition~\ref{prop:headroom}, a recommendation stops paying, though the two slopes are not the same object. The test was fixed before the first full run and the estimator could have passed it, which is what makes abstaining at this grain a measured requirement and not a taste for caution. It does not isolate design variance as the cause: a slope this low is also what a forecast shows when most of its variance is estimation noise, and the test cannot separate that from error in the approximate truth or from the mismatch between a four-week response and a $120$-week elasticity. Nor does it show that the gate selects the cells that forecast better: only $21$ UPCs pass it, and the one evaluable contrast, whether cells with more design variation forecast better, runs the other way ($-0.258$; $-0.509$ to $-0.045$). It does not evaluate Guards 1 to 8, and is reported as evidence about what the estimator forecasts on real data, not as validation of the system.

\subsection{What the governance costs against rules that do not abstain}\label{subsec:baselines}

This subsection compares the system's decisions, when the truth is known, with those a simpler rule would have taken, on the $200$ replications of Section~\ref{subsec:funnel} and on two variants of that generator that multiply every injected elasticity by $1.8$ (\texttt{high\_elasticity}) and by $2.6$ (\texttt{reversal}); a fourth variant, \texttt{cost\_link}, in which the competitor's price receives the cost shock that moves the instrument, serves the exclusion audit and the decomposition of Table~\ref{tab:decomp}, not the comparison of losses. A rule returns ACT or WAIT and, when it acts, takes the optimizer's price at its own estimate inside the corridor of $\pm10\%$ around the competitor's price. The \emph{decision loss} is the contribution of the better of holding and the optimum at the injected elasticity, minus the rule's, as a percentage of the contribution of holding; \emph{harm} is the share of evaluations in which the rule moved the price to a point worse than holding. Thirty applicable rules and four controls are compared (Supplementary Material, Table D.21; these rules are unrelated to the thirty pre-registered verification rules of Table D.2); Table~\ref{tab:other-market} and Supplementary Material, Table G.6 show the rules that matter, and intervals resample replications.

\paragraph{In the paper's generator, abstaining is expensive.} The true optimum raises the price in $99.4\%$ of the evaluations, so acting is almost always right whatever the estimate: never acting loses $17.1\%$, always acting $5.2\%$ ($4.7$ to $5.7$) and the system $15.0\%$ ($14.4$ to $15.7$), $2.1$ points less than never acting and $9.8$ points ($9.1$ to $10.4$) more than always acting. What the guards buy shows only in the harm rate, $0.002$ per evaluation for the system against $0.099$ for always acting and $0.016$ for a rule that acts at random as often as the system (Supplementary Material, Table F.2); this generator measures what abstention costs and cannot measure what it buys. Where acting can hurt, in \texttt{reversal}, always acting loses $12.7\%$ ($11.9$ to $13.6$) against $8.7\%$ ($7.9$ to $9.7$) for the system, and abstention is worth its cost; in \texttt{high\_elasticity} it is not ($7.5\%$ against $8.3\%$), although always acting harms $0.46$ of the evaluations.

\paragraph{Choosing among rules.} A mechanical criterion was fixed before the run: a rule is eligible if, in each of the three scenarios, it vetoes at most $0.20$ of the identifiable evaluations and acts on at most $0.05$ of the non-identifiable ones. One of the thirty applicable rules meets it (Table D.22), acting when the confidence interval is wholly negative and the partial $R^2$ of price exceeds $0.10$; its loss is lower than the system's in all three scenarios, by $2.6$ points ($2.2$ to $3.0$) in the paper's generator, $0.5$ in \texttt{high\_elasticity} and $0.6$ in \texttt{reversal}, while the system and every variant that keeps its guards fail the criterion on the false-veto limit, between $0.62$ and $0.80$ in the worst scenario. The winner uses the sign of the elasticity and a threshold chosen for this generator, and was selected among thirty rules on the same data that evaluate it, so it is a benchmark and not a validated replacement. On the measure the paper itself registered, the system as configured is not the best available rule.

\paragraph{The exclusion audit.} Guard~7 is the largest source of false vetoes, and the diagnostic evidence points to the nominal index that the cost series and the competitor's price share: run alone on panels where the restriction holds by construction, the audit vetoes $0.265$ of the identifiable presentations on the nominal series and $0.015$ once both series are divided by the generator's index, whereas with an injected violation it vetoes $0.438$ and $0.425$. We report this as a diagnostic: the deflated audit has not been run inside the full pipeline (Section~\ref{subsec:limitations}).

\subsection{Another market, the same decision function}\label{subsec:other-market}

The generators above belong to one family of designs, and the system's thresholds were set against it (Section~\ref{subsec:gate-generalization}). This subsection asks whether the cost and the benefit of abstaining survive a market written separately. The \emph{independent-market generator} shares with the others only the column schema of the panel: demand is logit-type, with an elasticity that rises with the price and is drawn for each presentation; list prices move through regional events, cost pass-through (so the cost instrument is relevant) and a lagged reaction to the competitor; the portfolio has ten presentations in four regions and $W=120$ weeks. A presentation is identifiable if none of four independently drawn failure mechanisms was assigned to it: a price that reacts to the previous week's demand shock (\emph{endogeneity}), little price variation, a price that follows the competitor's, and a month-end price cut that coincides with the pay-cycle surge in demand (\emph{calendar}); the label uses no statistic of the system and was fixed before any replication was run. Three scenarios differ in the median absolute elasticity: $2.4$ (the main scenario, fixed in advance, $150$ replications), $1.4$ and $3.6$ ($50$ each), on seeds disjoint from every other experiment. Rules, loss, harm and intervals are those of Section~\ref{subsec:baselines}, and Supplementary Material, Tables D.26 to D.34, give the protocol and every figure.

\begin{table}[htbp]
\centering\footnotesize\setlength{\tabcolsep}{3.5pt}
\caption{The cost of the decision function in the five-regime generator and in the independent market. \emph{False veto}: share of identifiable evaluations on which the rule waits; \emph{false ACT}: share of non-identifiable ones on which it acts. \emph{Two-statistic rule}: act when the confidence interval is wholly negative and the partial $R^2$ of price exceeds $0.10$, the rule that won in the five-regime generator. \emph{Loss}: decision loss in \% of the contribution of holding the price against the true optimum (lower is better); \emph{harm}: share of evaluations moved to a point worse than holding. Independent market: $150$ replications (base) and $50$ (inelastic, elastic); five-regime generator: $200$. Supplementary Material, Tables D.26 to D.31, give intervals and all rules.}\label{tab:other-market}
\resizebox{\linewidth}{!}{\begin{tabular}{@{}lccccccccccc@{}}
\toprule
 & \multicolumn{2}{c}{False veto} & \multicolumn{2}{c}{False ACT} & \multicolumn{4}{c}{Loss (\%)} & \multicolumn{2}{c}{Harm} \\
\cmidrule(lr){2-3}\cmidrule(lr){4-5}\cmidrule(lr){6-9}\cmidrule(l){10-11}
Generator, scenario & System & Gate + G1--4 & System & Two-stat. & Never & Always & Two-stat. & System & Always & System \\
\midrule
Five-regime, paper & $0.52$ & $0.15$ & $0.00$ & $0.00$ & $17.1$ & $5.2$ & $12.4$ & $15.0$ & $0.099$ & $0.002$ \\
Independent, base & $0.87$ & $0.74$ & $0.15$ & $0.70$ & $6.4$ & $2.7$ & $3.3$ & $5.6$ & $0.357$ & $0.036$ \\
Independent, inelastic & $0.81$ & $0.66$ & $0.16$ & $0.55$ & $12.3$ & $1.3$ & $6.4$ & $10.2$ & $0.164$ & $0.024$ \\
Independent, elastic & $0.91$ & $0.79$ & $0.09$ & $0.72$ & $7.6$ & $4.7$ & $3.7$ & $7.3$ & $0.412$ & $0.036$ \\
\bottomrule
\end{tabular}}
\end{table}

\paragraph{The false-veto rate is higher.} In the main scenario the system vetoes $0.874$ ($0.847$ to $0.900$) of the $602$ identifiable evaluations, against $0.522$ in the five-regime generator, and the gate with Guards 1 to 4 vetoes $0.744$ ($0.145$ there); the rate is $0.808$ and $0.907$ in the inelastic and elastic scenarios, and each of the three fails the limit of V9 by a wide margin, because the thresholds of Guards 4, 6 and 8 and of the width gate are absolute, calibrated to one scale (Section~\ref{subsec:limitations}).

\paragraph{What the system lets through.} Over the $898$ non-identifiable evaluations the system acts on $134$ ($0.149$; Wilson interval $0.127$ to $0.174$), against $0$ of $800$ in the five-regime generator. The failure is specific (Supplementary Material, Table G.7): three of the four mechanisms are caught (leaving out the panels with a price reaction to demand shocks, the false-ACT rate is $0.02$), but where the price reacts to the previous week's demand shock the system acts on $0.40$ ($0.33$ to $0.47$) of the evaluations, three times its rate on identifiable ones, with a bias of $+0.25$ in $\hat\theta$ and a coverage of $0.25$ of the true elasticity (Section~\ref{subsec:limitations}), because the reaction creates the price variation the gate rewards and the cost-instrument contrast of Guard~4 is too noisy to separate these panels from identifiable ones. The system therefore acts no more often on identifiable evaluations ($0.126$) than on non-identifiable ones ($0.149$).

\paragraph{Decision loss.} In the main scenario never acting loses $6.4\%$ of the contribution of holding, always acting $2.7\%$, the system $5.6\%$ ($5.2$ to $6.1$) and an oracle that abstains exactly where acting would hurt $0.8\%$; the system loses $2.9$ points ($2.4$ to $3.5$) more than always acting and $2.3$ ($1.8$ to $2.8$) more than the two-statistic rule, in the same order in the other scenarios. The system's loss is within $0.3$ points of that of a rule that acts at random at its own frequency ($5.9\%$), and most of the harm it avoids it avoids by acting rarely (Section~\ref{subsec:condition-check}).

\paragraph{Choosing among rules, and what this establishes.} No rule is eligible under the criterion of Section~\ref{subsec:baselines} in this generator (Table D.31): the two-statistic rule recommended there vetoes at most $0.30$ of the identifiable evaluations but acts on $0.55$ to $0.72$ of the non-identifiable ones, against $0.00$ there, so its advantage does not transfer, although the rules keep their order (Spearman correlation $0.93$ for the false-veto rate and $0.81$ for the false-ACT rate over the thirty non-oracle rules; Table D.33) and the finding on Guard~7 replicates (Table D.34). The cost of abstention measured in the five-regime generator is therefore not specific to it and is larger in a market written separately. This does not establish that abstaining is wrong: a well-aimed abstention has value (the oracle loses $0.8\%$ against $5.6\%$ for the system), but this system's is not well aimed in this market, and what it buys is the avoidance of harm, $0.04$ per evaluation against $0.36$ for always acting ($0.28$ for the two-statistic rule) and $0.05$ for a random rule that acts as often as the system, at the price of the contribution forgone. These are results about one more synthetic design, written by the author of the system, and are evidence about robustness to the form of the market, not about the channel.

\subsection{Where does abstaining pay? The condition against the records}\label{subsec:condition-check}

\paragraph{What was checked.} Propositions~\ref{prop:value} and~\ref{prop:headroom} were derived after the experiments of Sections~\ref{subsec:baselines} and~\ref{subsec:other-market}, not before them, and are checked here on the per-presentation records those experiments produced: seven scenarios and $4{,}960$ evaluations, with no new run. The slope $\hat\beta$ of a group is that of the log of the true optimum over the current price on the log of the recommended price over the current price, through the origin; the optimum is the corridor's, which understates the headroom wherever the corridor binds, and the moments are unweighted. The groups are those the generators define, the failure mechanism assigned to a presentation in the independent market and the regime in the five-regime generator: $38$ in all, none with fewer than $37$ evaluations, none defined by a statistic of the system.

\paragraph{The identity holds, and it says why the system loses to always acting.} Proposition~\ref{prop:value}(ii) holds to $10^{-12}$ in every scenario (Supplementary Material, Table F.1). Always acting beats the system exactly when the gain from acting on the presentations the system leaves waiting is positive (Proposition~\ref{prop:value}(i)), and it is, with an interval that excludes zero, in six of the seven scenarios: $3.4$ ($2.8$ to $4.0$) points in the main independent scenario, $11.6$ ($10.9$ to $12.4$) in the paper's generator and $0.86$ ($0.14$ to $1.63$) in \texttt{high\_elasticity}; in \texttt{reversal} it is $-4.3$ ($-5.3$ to $-3.3$), and the system beats always acting there, as Section~\ref{subsec:baselines} found.

\begin{figure}[htbp]
\centering
\includegraphics[width=0.82\linewidth]{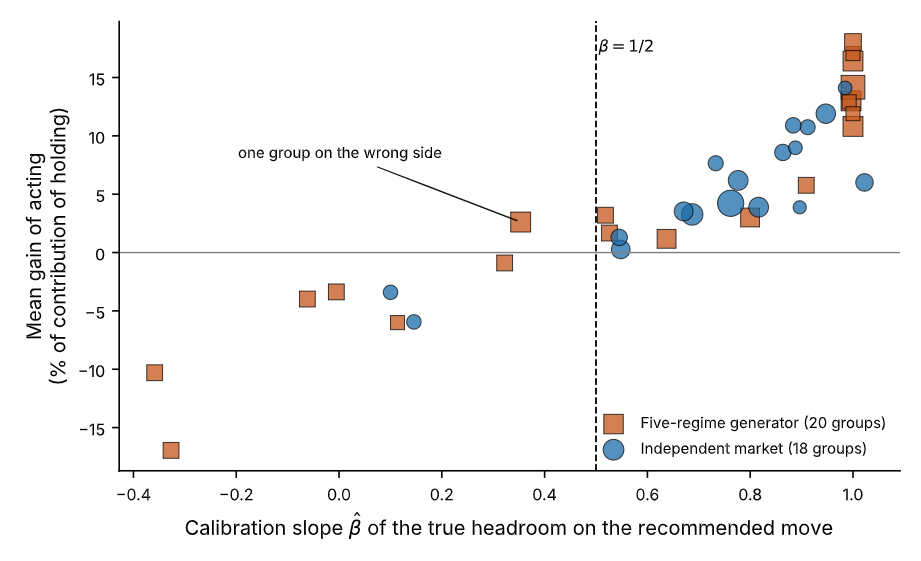}
\caption{The calibration slope and the gain from acting, $38$ groups defined by the generators (mechanism assigned or regime), by scenario. The condition of Proposition~\ref{prop:headroom} predicts a positive mean gain to the right of the dashed line and a negative one to its left. Marker area grows with the number of evaluations. The calibration slope $\hat\beta$ is that of the log of the true optimum over the current price on the log of the recommended price over the current price; the gain is in percent of the contribution of holding the price. Supplementary Material, Table F.3, lists the groups.}\label{fig:calibration}
\end{figure}

\paragraph{The condition locates the groups where acting hurts.} The slope $\hat\beta$ is below one half in nine of the $38$ groups, and the mean gain from acting is negative in eight of them (Figure~\ref{fig:calibration}). The sign of $2\hat\beta-1$ agrees with the sign of the mean gain in $37$ of the $38$ groups, against $30$ for a rule that predicts that acting pays everywhere, and the rank correlation between $\hat\beta$ and the mean gain is $0.92$. The exception, the round-price group of the paper's generator ($\hat\beta=0.35$, mean gain $2.6$), is one where the slope is below one half and acting still pays. The check is on generators written by the author, one of them independently of the system but not of its author, and is not evidence that the condition holds in a market.

\paragraph{Identification is not what separates the groups.} In the main independent scenario the six mechanism classes have $\hat\beta$ between $0.55$ and $1.02$, all above one half, including the presentations whose price reacts to the previous week's demand shock ($0.82$): acting gains $3.3\%$ on the non-identifiable presentations and $4.2\%$ on the identifiable ones, so by Corollary~\ref{cor:classes} no rule that works through the identifiable/non-identifiable classification can beat always acting there.

\begin{table}[htbp]
\centering\footnotesize\setlength{\tabcolsep}{3.5pt}
\caption{Where the system's abstentions pay. Groups are those the generators define; a group is \emph{below one half} if its calibration slope $\hat\beta$ is below $1/2$ (Figure~\ref{fig:calibration}). \emph{Share}: evaluations in groups below one half. \emph{ACT rate}: share of evaluations on which the system acts, within the groups below and not below one half. \emph{Value}: reduction in decision loss against always acting that the system's abstentions produce in each set of groups, in points of \% of the contribution of holding the price per evaluation (negative: the abstentions cost); the two parts add to the loss of always acting minus that of the system. Intervals resample replications ($2000$ draws); the grouping is fixed at its full-sample value. Supplementary Material, Tables F.1 to F.4, give the identities, the harm, all $38$ groups and the corridor.}\label{tab:decomp}
\resizebox{\linewidth}{!}{\begin{tabular}{@{}lccccccc@{}}
\toprule
 & Share & \multicolumn{2}{c}{ACT rate} & \multicolumn{3}{c}{Value of abstaining} \\
\cmidrule(lr){3-4}\cmidrule(lr){5-7}
Scenario & below one half & below & not below & below one half & not below one half & Total \\
\midrule
Independent market, main & 0.00 & --- & 0.140 & 0 & $-$2.93 [$-$3.47; $-$2.38] & $-$2.93 [$-$3.47; $-$2.38] \\
Independent market, inelastic & 0.00 & --- & 0.174 & 0 & $-$8.87 [$-$10.11; $-$7.70] & $-$8.87 [$-$10.11; $-$7.70] \\
Independent market, elastic & 0.22 & 0.000 & 0.121 & 1.03 [0.37; 1.68] & $-$3.63 [$-$5.07; $-$2.24] & $-$2.60 [$-$4.36; $-$1.07] \\
Five-regime, paper's generator & 0.17 & 0.000 & 0.191 & $-$0.44 [$-$1.04; 0.12] & $-$9.32 [$-$9.70; $-$8.93] & $-$9.75 [$-$10.43; $-$9.09] \\
Five-regime, cost link & 0.17 & 0.000 & 0.136 & 1.00 [0.14; 1.87] & $-$10.93 [$-$12.12; $-$9.74] & $-$9.93 [$-$11.54; $-$8.30] \\
Five-regime, high elasticity & 0.33 & 0.000 & 0.109 & 0.71 [0.01; 1.39] & $-$1.51 [$-$1.81; $-$1.22] & $-$0.80 [$-$1.52; $-$0.08] \\
Five-regime, reversal & 0.50 & 0.000 & 0.133 & 5.21 [5.00; 5.42] & $-$1.18 [$-$2.09; $-$0.19] & 4.03 [3.13; 5.01] \\
\bottomrule
\end{tabular}}
\end{table}

\paragraph{The system abstains where it should, and far beyond.} Table~\ref{tab:decomp} splits the value of the system's abstentions against always acting by whether the condition fails. In every scenario that has a group below one half the system never acts on those groups (ACT rate $0.000$), and those abstentions earn between $0.7$ and $1.0$ points in three scenarios and $5.2$ in \texttt{reversal}; in the paper's generator the one group below one half is the exception of the previous paragraph, and the same abstentions are worth $-0.4$ points. Elsewhere the system acts on $0.11$ to $0.19$ of the evaluations: it abstains on $81$ to $89$ percent of those where acting pays, at a cost of $1.2$ to $10.9$ points. Its shortfall against always acting is therefore not a failure to avoid the groups where acting hurts, as these generators define them; it is abstaining far beyond them. In the main independent scenario no group is below one half, and every abstention is a cost: $2.9$ ($2.4$ to $3.5$) points, the whole of the gap.

\paragraph{Abstention as a dial, not a verdict.} Against always acting the system forgoes $9.8$ points of contribution in the paper's generator and $2.9$ in the main independent scenario, and lowers the probability of a harmful move by $0.097$ and $0.321$; what that exchange is worth depends on the cost of a harmful move, which this paper does not set. Two facts bound the reading. Against a random rule that acts as often as the system, the harm reduction is clear in the five-regime generator ($0.002$ against $0.016$) and modest in the independent market ($0.036$ against $0.050$, with intervals that include zero in two of its three scenarios): most of the harm avoided is avoided by acting rarely. And the cost is not abstention aimed at the wrong groups, since the system never acts on the groups below one half, but abstaining on $81$ to $89$ percent of the evaluations where acting pays. The framework's contribution is that this exchange is measured and located by Propositions~\ref{prop:value} and~\ref{prop:headroom}, not that the present thresholds sit at a good point on it (Section~\ref{subsec:limitations}).

\paragraph{Harm and the corridor.} Proposition~\ref{prop:value} also applies to the probability of harm: in the main independent scenario always acting harms $0.36$ of the evaluations, the system $0.036$ ($0.027$ to $0.046$) and a random rule that acts at the system's frequency $0.050$ ($0.043$ to $0.057$) (Supplementary Material, Table F.2). The ceiling of the corridor is the true optimum in $0.39$ of the evaluations of the main independent scenario, but the optimum is a bound of the corridor, the floor included, in $0.79$, $0.81$ and $0.90$ of the evaluations of the three scenarios (Table F.4), so by Proposition~\ref{prop:direction} the direction of the recommended move decides the sign of the gain.

\paragraph{What this establishes and what it does not.} It establishes an exact accounting of what an abstention rule can earn (Proposition~\ref{prop:value}), a condition under which abstaining pays that predicts the sign of the gain in $37$ of $38$ groups of two families of generators, and a diagnosis of the system against it: it abstains where the condition fails and on most evaluations where it holds. It does not show how to estimate $\beta$ without the true optimum, and it does not test the condition on real prices: the optimum is unobserved there, and the calibration slope of $0.22$ of Section~\ref{subsec:real-data} is that of the response to price, not of the recommended move.

\section{Operational Governance and Conclusion}\label{sec:governance}

\subsection{Where the value of a system like this one lies}\label{subsec:contribution}

In intermediated retail the binding constraint is not the optimizer: the input slope is the fragile link, and optimization is the \emph{simplest} component of the system. Nor does value concentrate in the estimator, since which of two nuisance learners is better reverses between two generators. What survives is narrower and more durable: the value lies in \emph{measuring which of the system's own claims hold}, and in publishing the measurement; Section~\ref{subsec:baselines} turns that discipline on the system itself, since its false-veto rate fails the paper's own limit and a simpler rule has lower decision loss in every scenario tried. The three contributions are complementary: causal elasticity without an admissibility gate cannot say when to trust itself, a gate without the aggregation result enforces a precision standard at a unit the evidence cannot reach, and either without the tacit-assumption audit is blind to failure modes that are properties of the environment rather than of the estimator.

\subsection{Governance, adoption and the recalibration this run forces}\label{subsec:artifact}

The artifacts feed a human-in-the-loop process, not an autopilot: a periodic trigger, the recommendation layer, commercial validation with a documented override, a committee, and execution with ex-post measurement. The harder conversation for a committee is that the system cannot, on the evidence this channel supplies, publish a per-presentation elasticity with an interval narrow enough to act on, and here is the level at which it can. An engine designed to halt rather than degrade returns WAIT for a substantial share of a real portfolio, which a committee expecting a price list reads as a malfunction; the countermeasure to the predictable pressure to relax a threshold or publish the midpoint is an explicit degraded mode (no recommendation, reason, date of next review) and the decision arithmetic with every abstention.

The run adds a second governance item. The admissibility width threshold of $0.6$ and the construction of the interval it reads were set independently, and Section~\ref{subsec:impossibility} reports that the interval this implementation currently produces does not reliably clear that threshold at the presentation level. It is one decision with three defensible resolutions, each to be presented costed: publish at a coarser unit, where the interval is narrowest; publish per presentation with an interval that does not cover, and say so on the artifact; or supplement the observational panel with experimental variation, a path Supplementary Material, Appendix D takes up empirically and companion methodological work \citep{delgado2026acrossdesignuncertaintyshortpricing} develops further. The first is available now; the second is what the system does today without declaring it.

A WAIT verdict is itself diagnostic. When the admissibility gate fails for want of price variation, or Guard~2's robustness value flags that an unmeasured confounder could overturn the estimate, the shortfall is one of identification, and only independently assigned variation resolves it; when the trigger is Guard~8's quota-pressure bracket or the contamination test, the threat is operational, and the finding is a requirement on how an experiment must be run, with incentive discretion frozen and treatment and control drawn at a geographic scale wide enough that a competitor's reaction cannot cross between them.

The ex-post loop remains the most valuable asset: the residual uncertainty this validation isolates behaves like dispersion across designs rather than sampling noise within one, so better estimation of the same panel narrows it only partially, while generating new designs narrows it further. For a pilot, should one be run (none is part of this study), success criteria are stated in advance rather than reported as results, and the $2$--$5\%$ revenue target is a commercial ambition and not a forecast the evidence supports \citep{marn1992managing, deng2023alibaba, llenas2026pepsico}; a holdout should be sized against the experimental layer's own measured resolution, between twelve and thirty independently assigned units in this panel (Supplementary Material, Table D.13).

\subsection{Declared limitations}\label{subsec:limitations}

This section consolidates the limitations that Section~\ref{sec:results} and the Supplementary Material flag individually. The cross elasticity inherits the endogeneity of the competitor's price, for which no instrument exists with these data; the competitive scenarios are reduced-form, one round, with a constant reaction; and the performativity of the engine has no econometric fix, only an experimental one.

Two limitations attach to the classification of Definition~\ref{def:extended-classification} and are properties of the evidence rather than of the rule. First, the \textsc{transient} category is empirically empty: Guard~5 is not operative in the validated configuration (Supplementary Material, Appendix D) and binds in $0.0\%$ of the $108$ instances of the extended-taxonomy study (Section~\ref{subsec:extended-taxonomy}). A reader is entitled to treat the third branch as untested, and the identification branch, the only one that hands a presentation to an experiment, is thin as well ($2$ of the $89$ blocked instances). Second, the funnel of Section~\ref{subsec:funnel} reports per-cause incidence for six of the ten causes (Guard~1's two causes jointly) and folds the other four into the ACT row, because the aggregator does not record separate exits for Guards~5 through~8; Section~\ref{subsec:extended-taxonomy} and Supplementary Material, Table D.24 ($200$ replications, $1\,200$ evaluations) resolve all ten causes on other populations, complementing the funnel's figures.

Eleven further limitations were produced by the validation run itself, and each qualifies a claim made elsewhere in this paper.

\begin{enumerate}\tightlist
  \item \textbf{The interval this implementation currently produces falls short of nominal coverage at the presentation level, and the shortfall is one of centering rather than width.} The Paule--Mandel construction that reaches nominal coverage (Section~\ref{subsec:hierarchical-coverage}) is \emph{not yet adopted}: in an illustrative run (seed $1000$, six presentations) it would have cut the presentations released as \textsc{act} from two to zero and widened the mean interval from $3.68$ to $11.70$. No figure in this paper uses it.
  \item \textbf{No aggregation level tried both covers and clears the admissibility gate at its current threshold.} In none of the twelve level-learner-scenario combinations does an interval with coverage of at least $0.90$ fit inside the threshold; those that reach it are $1.55$ to $3.38$ times as wide (Table D.15). On public scanner data the category-level interval clears the gate with no showing that it covers a future window (Section~\ref{subsec:real-data}).
  \item \textbf{Guard~5 is not operative.} The break tests reject at $0.94$ under no break with the asymptotic critical value; with a calibrated critical value the size is at most $0.107$, but power at the largest break injected does not exceed $0.161$ ($0.286$ inside the recency window), against a pre-specified $\ge 0.50$ (Table D.7; rule V6 of Table D.2). The recency veto and the sequential reset are withdrawn, and the currency of the elasticity rests on the drift alert and the revalidation cadence.
  \item \textbf{The gate is calibrated, not general.} Outside the configuration the frozen thresholds were calibrated on, precision ranges from $0.000$ to $0.75$ (undefined where the gate admits nothing), against a rule of $\ge 0.80$, and on the calibration configuration itself it is $0.5$ (Section~\ref{subsec:gate-generalization}). The thresholds are absolute, so in the independent market the false-veto rate of the full decision function rises from $0.81$ to $0.87$ and $0.91$ as the median absolute elasticity goes from $1.4$ to $2.4$ and $3.6$ (Section~\ref{subsec:other-market}).
  \item \textbf{The Dual Beta, the estimator that separates the elasticity to price increases from that to price cuts, is withdrawn.} Size $0.286$ against power $0.321$ on the nominal specification, $0.321$ against $0.357$ on the deflated one; the lookup table loses its per-direction band.
  \item \textbf{The conformal guarantee is calibrated within the historical regime.} A band can be narrow and wrong if the regime shifts, and the defense against that lived in the structural-break guard, which this run reports as not operative.
  \item \textbf{The exclusion audit (Guard~7) is the largest single source of false vetoes, and as run it measures the shared price index as much as the exclusion restriction.} It vetoes $0.33$ of identifiable presentations where the restriction holds by construction, against $0.47$ under an injected violation; dividing cost and competitor price by the common index lowers the first from $0.265$ to $0.015$ and the second only from $0.438$ to $0.425$ (Section~\ref{subsec:baselines}; different panels from the full pipeline, in which the deflated audit has not been run).
  \item \textbf{The system as configured is not the best available rule on the measure the paper itself registered.} Its false-veto rate on identifiable presentations is $0.52$ against a limit of $0.20$ in the five-regime generator and $0.87$ in an independently written market, where always acting also loses less than the system in all three scenarios; a two-statistic rule has lower decision loss in every scenario of both, although in the independent market it acts on $0.55$ to $0.72$ of the non-identifiable presentations (Sections~\ref{subsec:baselines} and~\ref{subsec:other-market}). The thresholds were not retuned: the system is reported as validated, not as optimal.
  \item \textbf{Price reaction to demand shocks is not detected.} Where the list price reacts to the previous week's demand shock, the system acts on $0.40$ of the evaluations, three times its rate on identifiable ones, with a bias of $+0.25$ in $\hat\theta$ and a coverage of $0.25$; Guard~4, built to read residual endogeneity, fires on $0.15$ of the identifiable ones, and no guard tests whether the price depends on past demand (Section~\ref{subsec:other-market}).
  \item \textbf{The forecasts of single-window estimates are poor on real data, and the gate could not be tested there.} The estimate of one window forecasts the response in the next with a calibration slope of $0.22$ and a skill of $-0.25$; the gate passes $21$ UPCs with events and the variant that counts between-window variance passes none, so whether the gate selects better-forecasting cells could not be tested, and post hoc variants of the decision advantage do not favour the estimate over the category median (Supplementary Material, Table D.39). The approximate truth is not an experiment and the Guards are not evaluated (Section~\ref{subsec:real-data}).
  \item \textbf{The condition for abstaining to pay is derived in a local model and cannot be checked on real prices.} It assumes a quadratic contribution near the optimum, was derived after the experiments of Sections~\ref{subsec:baselines} and~\ref{subsec:other-market}, and is checked only on the records they produced (Section~\ref{subsec:condition-check}); its slope needs the true optimum, which no real panel records. A rule built to target the condition has not been designed or tested here.
\end{enumerate}

Two further items run against the system: a boosted nuisance learner is not dominated by a regularized one on the extended generator, and the round-point failure does not reproduce on the experiment built to demonstrate it. The cross-fitting partition is unembargoed (Supplementary Material, Appendix A.1), so a week can appear on both sides of a fold boundary through a different region; the embargoed variant was measured but not adopted, to preserve comparability across the ablation battery (Appendix D), so every figure in this paper carries a declared and unquantified leak.

\paragraph{Limitations of scope.} The system has not been run on a commercial panel, and no pilot is part of this study, so nothing reported here measures commercial uplift; the $2$--$5\%$ figure of Section~\ref{subsec:artifact} is a target and not an estimate. The two real-data checks run only the estimation-and-interval core on a public supermarket panel, and neither evaluates Guards 1 to 8, recovery of a known parameter, or the intermediated channel (Section~\ref{subsec:real-data}). Both generators are stationary apart from an inflationary drift and inject their confounding at magnitudes the designer chose, so agreement between them is evidence of robustness to configuration and not of external validity. The measurement of what the governance costs and buys rests on two synthetic families written by the author of the system: in the first, in its own configuration, the right move is almost always to raise the price, and the two stress variants change the elasticity only by a common factor; the second varies the form of the market but is one design. Together they show the direction of the trade-off between abstaining and acting and not its size on a real category. Promotional incrementality is outside the system's scope. Every governance threshold must be recalibrated against a real category before being fixed (Supplementary Material, Appendix E and Table D.14).

\paragraph{Evidence without confidential data.} Panels of intermediated retail channels are commercial assets held under non-disclosure agreements and are not publicly accessible; outside readers cannot check a result computed on one, so the paper reports none, and its evidence rests on two sources that independent researchers can examine: a transparent synthetic environment with known ground truth, whose generator is public \citep{pricingpanel}, and the public Dominick's Finer Foods scanner database \citep{kilts_dominicks} (Section~\ref{subsec:real-data}). What this design cannot show, and the paper does not claim, is recovery on the intermediated channel itself.

\subsection{Conclusion}\label{sec:conclusion}

This paper asked, of an operationally complete causal pricing system, not what the elasticity is but when there is enough evidence to act on it and when honest conduct is to wait. The system was validated against two data-generating processes with known ground truth under thirty pre-registered rules, its decision function was measured in a third, independently written market, and its estimates were compared with real price changes in public scanner data. Ten rules pass, nineteen fail, and one does not apply; that the failures outnumber the passes is the result rather than a caveat on it, since the value of this validation lies in the places where the system's own claims did not survive the test written for them.

Each of the three contributions answers one question posed in Section~\ref{sec:intro}. The ACT/WAIT/EXPERIMENT framework (Section~\ref{subsec:decision-layer}) answers \emph{when one can act}: a gate that abstains rather than publish a number the evidence does not support, at a measured false-veto cost of $0.52$ on presentations that are identifiable by construction ($0.15$ for the gate with Guards 1 to 4; Section~\ref{subsec:funnel}), with thresholds calibrated on one configuration and not shown to be general. Each of the ten abstention causes maps to exactly one of three verdict types, and \textsc{wait-identification} is a formal statement of experimental eligibility; this paper does not design that experiment. The conservatism is priced in decisions as well (Sections~\ref{subsec:baselines} and~\ref{subsec:other-market}): in the paper's generator, where acting is almost always right, the system leaves $15.0\%$ of the contribution on the table against $5.2\%$ for always acting, and a two-statistic rule loses less than the system in every scenario tried; in the independently written market its false-veto rate is $0.87$, and it acts on $0.15$ of the non-identifiable presentations and buys a harm rate of $0.04$ per evaluation against $0.36$ for always acting. In the validation generators it abstains in every group in which the condition for acting to pay fails ($9$ of $38$) and, in addition, on $81$ to $89$ percent of the evaluations in the groups where it holds (Section~\ref{subsec:condition-check}); the condition cannot be checked on real prices, where the estimates forecast poorly at the level of a single UPC (Section~\ref{subsec:real-data}).

The admissibility audit (Section~\ref{subsec:tacit}) answers \emph{what has to hold for that gate to be trusted}: seven assumptions a competent build otherwise leaves tacit, converted into executable contrasts, four computed and published by the system before it recommends and three run as validation audits against a known truth. The aggregation result (Section~\ref{subsec:unit-of-identification}) answers \emph{at what level a decision is legitimate}: the root mean squared error falls from $0.488$ at the presentation level to $0.159$ at the category level under the sieve learner, and from $0.571$ to $0.336$ under the boosted learner, and on public scanner data the same procedure points to the category, where the hierarchical interval clears the width gate in $7$ of $11$ and $9$ of $11$ categories and almost nowhere below (Section~\ref{subsec:real-data}). If the dispersion that this run isolates behaves like variation \emph{across} designs rather than noise within one, better estimation of the same panel would narrow it only partially (Section~\ref{subsec:limitations}). This paper does not resolve that limitation.

What it delivers instead are the three contributions above, and a system that would rather report a coarser unit, an interval that declares what it does and does not cover, and an abstention whose cost has been measured, than publish a number it cannot support. For a reader building or buying such a system, the transferable rule is to judge an abstention rule by the calibration of what it lets through, to publish at the coarsest unit the evidence supports, and to price every abstention in contribution and in harm before relaxing a threshold.

\section*{Declarations}\label{sec:declarations}
\addcontentsline{toc}{section}{Declarations}

\paragraph{Data availability.} The results reported in this paper are produced by a synthetic data-generating process that contains no commercial information; it is described in Supplementary Material, Appendix E (The Data-Generating Process), and its generator is public \citep{pricingpanel}. The replication code for the independent-market generator (Section~\ref{subsec:other-market}) and for the decision validation on Dominick's data (Section~\ref{subsec:real-data}) accompanies the Supplementary Material. The exception to the synthetic data is the two checks of Section~\ref{subsec:real-data}, which use the public Dominick's Finer Foods database \citep{kilts_dominicks}, provided for academic research by the Kilts Center for Marketing at the University of Chicago Booth School of Business and not redistributed here (Supplementary Material, Table D.16 identifies the file versions). The system is designed to consume commercial panels, which are proprietary and not publicly available; none was used to produce any result reported here (Section~\ref{subsec:limitations}).

\paragraph{Funding.} This research received no financial support from any public, commercial, or non-profit organization. The work was conducted independently, without institutional funding or the use of proprietary data.

\paragraph{Conflicts of interest.} The author declares no competing interests. No proprietary or commercial data were used in this study.

\paragraph{Author contributions.} \textbf{Pedro Cadahia Delgado:} Conceptualization, Methodology, Software, Validation, Formal analysis, Investigation, Data curation, Original draft preparation, Review and editing, Visualization.

\paragraph{Ethics approval.} This study involves no human or animal subjects.
It analyses a synthetic generator and public store-level scanner data aggregated by week, and no
personal data are processed at any stage.

\bibliographystyle{plainnat}
\bibliography{references}

@article{chernozhukov2018double,
  title     = {Double/debiased machine learning for treatment and structural parameters},
  author    = {Chernozhukov, Victor and Chetverikov, Denis and Demirer, Mert and Duflo, Esther and Hansen, Christian and Newey, Whitney and Robins, James},
  journal   = {The Econometrics Journal},
  volume    = {21},
  number    = {1},
  pages     = {C1--C68},
  year      = {2018},
  publisher = {Oxford University Press}
}

@article{rubin1980randomization,
  title     = {Randomization analysis of experimental data: the {F}isher randomization test comment},
  author    = {Rubin, Donald B.},
  journal   = {Journal of the American Statistical Association},
  volume    = {75},
  number    = {371},
  pages     = {591--593},
  year      = {1980}
}

@article{lucas1976econometric,
  title     = {Econometric policy evaluation: a critique},
  author    = {Lucas, Robert E.},
  journal   = {Carnegie-Rochester Conference Series on Public Policy},
  volume    = {1},
  pages     = {19--46},
  year      = {1976}
}

@book{mackenzie2006engine,
  title     = {An Engine, Not a Camera: How Financial Models Shape Markets},
  author    = {MacKenzie, Donald},
  publisher = {MIT Press},
  address   = {Cambridge, MA},
  year      = {2006}
}

@article{kunsch1989jackknife,
  title     = {The jackknife and the bootstrap for general stationary observations},
  author    = {K{\"u}nsch, Hans R.},
  journal   = {Annals of Statistics},
  volume    = {17},
  number    = {3},
  pages     = {1217--1241},
  year      = {1989}
}

@article{paule1982consensus,
  title     = {Consensus values and weighting factors},
  author    = {Paule, Robert C. and Mandel, John},
  journal   = {Journal of Research of the National Bureau of Standards},
  volume    = {87},
  number    = {5},
  pages     = {377--385},
  year      = {1982}
}

@article{besanko2005own,
  title     = {Own-brand and cross-brand retail pass-through},
  author    = {Besanko, David and Dub{\'e}, Jean-Pierre and Gupta, Sachin},
  journal   = {Marketing Science},
  volume    = {24},
  number    = {1},
  pages     = {123--137},
  year      = {2005},
  doi       = {10.1287/mksc.1030.0043}
}

@article{levy2011price,
  title     = {Price points and price rigidity},
  author    = {Levy, Daniel and Lee, Dongwon and Chen, Haipeng and Kauffman, Robert J. and Bergen, Mark},
  journal   = {The Review of Economics and Statistics},
  volume    = {93},
  number    = {4},
  pages     = {1417--1431},
  year      = {2011}
}

@article{abadie2010synthetic,
  title     = {Synthetic control methods for comparative case studies: estimating the effect of {C}alifornia's tobacco control program},
  author    = {Abadie, Alberto and Diamond, Alexis and Hainmueller, Jens},
  journal   = {Journal of the American Statistical Association},
  volume    = {105},
  number    = {490},
  pages     = {493--505},
  year      = {2010}
}

@book{talluri2006theory,
  title     = {The Theory and Practice of Revenue Management},
  author    = {Talluri, Kalyan T. and van Ryzin, Garrett J.},
  publisher = {Springer},
  address   = {New York},
  year      = {2006}
}

@article{marn1992managing,
  title     = {Managing price, gaining profit},
  author    = {Marn, Michael V. and Rosiello, Robert L.},
  journal   = {Harvard Business Review},
  volume    = {70},
  number    = {5},
  pages     = {84--94},
  year      = {1992}
}

@article{hormby2010marriott,
  title     = {{M}arriott {I}nternational increases revenue by implementing a group pricing optimizer},
  author    = {Hormby, Sharon and Morrison, Julia and Dave, Prashant and Meyers, Michele and Tenca, Tim},
  journal   = {Interfaces},
  volume    = {40},
  number    = {1},
  pages     = {47--57},
  year      = {2010},
  doi       = {10.1287/inte.1090.0482}
}

@article{deng2023alibaba,
  title     = {{A}libaba realizes millions in cost savings through integrated demand forecasting, inventory management, price optimization, and product recommendations},
  author    = {Deng, Yuming and Zhang, Xinhui and Wang, Tong and Wang, Lin and Zhang, Yidong and Wang, Xiaoqing and Zhao, Su and Qi, Yunwei and Yang, Guangyao and Peng, Xuezheng},
  journal   = {INFORMS Journal on Applied Analytics},
  volume    = {53},
  number    = {1},
  pages     = {32--46},
  year      = {2023},
  doi       = {10.1287/inte.2022.1145}
}

@article{llenas2026pepsico,
  title     = {{P}epsi{C}o deploys {AI}-driven pricing and promotion optimization at scale},
  author    = {Llenas, Aleix and Salazar-Trevi{\~n}o, Eduardo and Leskovar, Francisco and Pons-Llopis, Francesc and Todeschini, Federico and Gowda, Deepika and Bofill, David and Anish, Laxmi and Cleavinger, Michael},
  journal   = {INFORMS Journal on Applied Analytics},
  year      = {2026},
  note      = {Published online June 11, 2026},
  doi       = {10.1287/inte.2025.0302}
}

@article{dietvorst2015algorithm,
  title     = {Algorithm aversion: people erroneously avoid algorithms after seeing them err},
  author    = {Dietvorst, Berkeley J. and Simmons, Joseph P. and Massey, Cade},
  journal   = {Journal of Experimental Psychology: General},
  volume    = {144},
  number    = {1},
  pages     = {114--126},
  year      = {2015}
}

@article{dietvorst2018overcoming,
  title     = {Overcoming algorithm aversion: people will use imperfect algorithms if they can (even slightly) modify them},
  author    = {Dietvorst, Berkeley J. and Simmons, Joseph P. and Massey, Cade},
  journal   = {Management Science},
  volume    = {64},
  number    = {3},
  pages     = {1155--1170},
  year      = {2018}
}

@article{berry1994estimating,
  title={Estimating discrete-choice models of product differentiation},
  author={Berry, Steven T.},
  journal={The RAND Journal of Economics},
  volume={25},
  number={2},
  pages={242--262},
  year={1994}
}

@article{berry1995automobile,
  title={Automobile prices in market equilibrium},
  author={Berry, Steven and Levinsohn, James and Pakes, Ariel},
  journal={Econometrica},
  volume={63},
  number={4},
  pages={841--890},
  year={1995}
}

@article{nevo2001measuring,
  title={Measuring market power in the ready-to-eat cereal industry},
  author={Nevo, Aviv},
  journal={Econometrica},
  volume={69},
  number={2},
  pages={307--342},
  year={2001}
}

@article{villasboas2007vertical,
  title={Vertical relationships between manufacturers and retailers: Inference with limited data},
  author={Villas-Boas, Sofia Berto},
  journal={The Review of Economic Studies},
  volume={74},
  number={2},
  pages={625--652},
  year={2007}
}

@article{nakamura2010accounting,
  title={Accounting for incomplete pass-through},
  author={Nakamura, Emi and Zerom, Dawit},
  journal={The Review of Economic Studies},
  volume={77},
  number={3},
  pages={1192--1230},
  year={2010}
}

@inproceedings{gibbs2021adaptive,
  title={Adaptive conformal inference under distribution shift},
  author={Gibbs, Isaac and Cand{\`e}s, Emmanuel},
  booktitle={Advances in Neural Information Processing Systems},
  volume={34},
  pages={1660--1672},
  year={2021}
}

@article{barber2023conformal,
  title={Conformal prediction beyond exchangeability},
  author={Barber, Rina Foygel and Cand{\`e}s, Emmanuel J. and Ramdas, Aaditya and Tibshirani, Ryan J.},
  journal={The Annals of Statistics},
  volume={51},
  number={2},
  pages={816--845},
  year={2023}
}

@article{elyaniv2010foundations,
  title={On the foundations of noise-free selective classification},
  author={El-Yaniv, Ran and Wiener, Yair},
  journal={Journal of Machine Learning Research},
  volume={11},
  pages={1605--1641},
  year={2010}
}

@inproceedings{mozannar2020consistent,
  title={Consistent estimators for learning to defer to an expert},
  author={Mozannar, Hussein and Sontag, David},
  booktitle={Proceedings of the 37th International Conference on Machine Learning},
  pages={7076--7087},
  year={2020}
}

@article{chiang2022multiway,
  title={Multiway cluster robust double/debiased machine learning},
  author={Chiang, Harold D. and Kato, Kengo and Ma, Yukun and Sasaki, Yuya},
  journal={Journal of Business \& Economic Statistics},
  volume={40},
  number={3},
  pages={1046--1056},
  year={2022}
}

@inproceedings{perdomo2020performative,
  title={Performative prediction},
  author={Perdomo, Juan and Zrnic, Tijana and Mendler-D{\"u}nner, Celestine and Hardt, Moritz},
  booktitle={Proceedings of the 37th International Conference on Machine Learning},
  pages={7599--7609},
  year={2020}
}

@article{villas1999endogeneity,
  title={Endogeneity in brand choice models},
  author={Villas-Boas, J Miguel and Winer, Russell S},
  journal={Management science},
  volume={45},
  number={10},
  pages={1324--1338},
  year={1999},
  publisher={INFORMS}
}

@book{pearl2009causality,
  title={Causality},
  author={Pearl, Judea},
  year={2009},
  publisher={Cambridge university press}
}

@misc{pricingpanel,
author = {Cadahia, Pedro},
license = {Apache-2.0},
month = aug,
title = {{pricing-experiments-synthetic-panel: Synthetic Panel Data Generator for Pricing Experiments}},
url = {https://github.com/pedrocadahia/pricing-experiments-synthetic-panel},
version = {0.1.0},
year = {2026}
}

@misc{delgado2026acrossdesignuncertaintyshortpricing,
      title={Across-Design Uncertainty in Short Pricing Panels: inference and Identification},
      author={Pedro Cadahia Delgado},
      year={2026},
      eprint={2608.21334},
      archivePrefix={arXiv},
      primaryClass={cs.LG},
      url={https://arxiv.org/abs/2608.21334}
}

@misc{kilts_dominicks,
  title        = {{Dominick's Finer Foods Database}},
  author       = {{Kilts Center for Marketing}},
  year         = {n.d.},
  howpublished = {University of Chicago Booth School of Business. \url{https://www.chicagobooth.edu/research/kilts/research-data/dominicks}},
  note         = {Accessed October 2026}
}

@incollection{mincer1969evaluation,
  author    = {Mincer, Jacob A. and Zarnowitz, Victor},
  title     = {The Evaluation of Economic Forecasts},
  booktitle = {Economic Forecasts and Expectations: Analysis of Forecasting Behavior and Performance},
  editor    = {Mincer, Jacob A.},
  publisher = {National Bureau of Economic Research},
  year      = {1969},
  pages     = {3--46}
}

@article{campbell2008predicting,
  author  = {Campbell, John Y. and Thompson, Samuel B.},
  title   = {Predicting Excess Stock Returns Out of Sample: Can Anything Beat the Historical Average?},
  journal = {Review of Financial Studies},
  volume  = {21},
  number  = {4},
  pages   = {1509--1531},
  year    = {2008}
}

@article{abadie2020sampling,
  author  = {Abadie, Alberto and Athey, Susan and Imbens, Guido W. and Wooldridge, Jeffrey M.},
  title   = {Sampling-Based versus Design-Based Uncertainty in Regression Analysis},
  journal = {Econometrica},
  volume  = {88},
  number  = {1},
  pages   = {265--296},
  year    = {2020},
  doi     = {10.3982/ECTA12675}
}

@article{manski2004statistical,
  author  = {Manski, Charles F.},
  title   = {Statistical Treatment Rules for Heterogeneous Populations},
  journal = {Econometrica},
  volume  = {72},
  number  = {4},
  pages   = {1221--1246},
  year    = {2004}
}

@article{rossi2014even,
  author  = {Rossi, Peter E.},
  title   = {Even the Rich Can Make Themselves Poor: A Critical Examination of {IV} Methods in Marketing Applications},
  journal = {Marketing Science},
  volume  = {33},
  number  = {5},
  pages   = {655--672},
  year    = {2014}
}

\end{document}